\documentclass[aoas,preprint]{imsart}

\RequirePackage{amsthm,amsmath,amsfonts,amssymb}
\RequirePackage[authoryear]{natbib}
\RequirePackage[colorlinks,citecolor=blue,urlcolor=blue]{hyperref}
\RequirePackage{graphicx}

\usepackage{bbm}

\startlocaldefs

\theoremstyle{plain}

\newtheorem{theorem}{Theorem}[section]

\newtheorem{remark}[theorem]{Remark}
\newtheorem{condition}[theorem]{Condition}

\theoremstyle{definition}

\def\ep{\textsf{E}}
\def\var{\textsf{Var}}       

\endlocaldefs

\begin{document}

\begin{frontmatter}
\title{A robust, scalable K-statistic for quantifying immune cell clustering in spatial proteomics data}
\runtitle{A robust, scalable K-statistic for quantifying immune cell clustering}

\begin{aug}
\author[A]{\fnms{Julia}~\snm{Wrobel}\ead[label=e1]{julia.wrobel@emory.edu}}
\and
\author[B]{\fnms{Hoseung}~\snm{Song}\ead[label=e2]{hoseung@kaist.ac.kr}}
\address[A]{Department of Biostatistics and Bioinformatics,
Emory University\printead[presep={ ,\ }]{e1}}

\address[B]{Department of Industrial and Systems Engineering,
KAIST \printead[presep={,\ }]{e2}}
\end{aug}


\begin{abstract}
Spatial summary statistics based on point process theory are widely used to quantify the spatial organization of cell populations in single-cell spatial proteomics data. Among these, Ripley's $K$ is a popular metric for assessing whether cells are spatially clustered or are randomly dispersed. However, the key assumption of spatial homogeneity is frequently violated in spatial proteomics data, leading to overestimates of cell clustering and colocalization. To address this, we propose a novel method, termed \textit{KAMP} (\textbf{K} adjustment by \textbf{A}nalytical \textbf{M}oments of the \textbf{P}ermutation distribution), for quantifying the spatial organization of cells in spatial proteomics samples. \textit{KAMP} leverages background cells in each sample along with a new closed-form representation of the first and second moments of the permutation null distribution of Ripley's $K$. Our method is robust to inhomogeneity, computationally efficient even in large datasets, and provides approximate $p$-values to test spatial clustering and colocalization. Methodological developments are motivated by a spatial proteomics study of women with ovarian cancer; in the subset with sufficient B cells and macrophages, \textit{KAMP} provides exploratory, scale-specific evidence linking B cell–macrophage colocalization with overall patient survival. Notably, we also find evidence that using $K$ without correcting for sample inhomogeneity may bias hazard ratio estimates in downstream analyses. 
\end{abstract}


\end{frontmatter}


\section{Introduction} \label{sec:intro}

Recent advances in single-cell spatial proteomics technologies have allowed researchers to examine tissue structure and function at the cellular level while preserving the original spatial context of the sample \citep{vandereyken2023methods, wrobel2023statistical}. These technologies facilitate the measurement of protein abundance \textit{in situ}, offering the potential to uncover novel spatial relationships between different cell types and their implications for patients \citep{bressan2023dawn}. Each patient has one or multiple tissue samples that are placed on a slide and imaged, cells are segmented and labeled, typically via semi-automated cell phenotyping algorithms, and then the spatial relationship between cells within and across samples is analyzed. Due to the lack of anatomical correspondence across samples, it is necessary to select summary features that quantify spatial relationships among cell types of interest. How to best summarize spatial features remains an active area of research.

It is common to represent cell counts and locations in each sample as one realization of a point process, called a ``point pattern''. A frequently employed method for analyzing spatial point patterns is the homogeneous point process model, which assesses whether cells are randomly distributed within the sample. To this end, Ripley's $K$ \citep{ripley1988statistical}, a popular metric of point process analysis, has been used to summarize the spatial organization of cells in proteomics data \citep{wilson2021challenges}. Specifically, $K$ has been used to demonstrate that the degree of clustering of various immune cell types is significantly associated with survival in ovarian and breast cancers \citep{keren2018structured, wilson2022tumor}. 

In this framework, $K$ refers to Ripley's $K$ function, a cumulative spatial summary that, at radius $r$, measures the expected number of target cells within distance $r$ of a randomly chosen cell of the same type after normalizing for cell density. For sample $s$, the observed value is denoted $\hat{K}_s(r)$ and is compared to the expected value under a null hypothesis of no spatial clustering, denoted $K_0(r)$. When $\hat{K}_s(r) > K_0(r)$, spatial clustering is inferred, with $\tilde{K}_s(r) = \hat{K}_s(r) - K_0(r)$ quantifying the ``degree of clustering''. This quantity is then used as a covariate in models of patient outcomes. A bivariate form of Ripley's $K$ also exists and can be used to quantify spatial colocalization between two point types. In the context of spatial proteomics, this bivariate statistic often measures the spatial interaction between two immune cell types. Except where otherwise noted, we use the terms clustering and colocalization interchangeably.

Under the standard $\hat{K}_s(r)$ assumption of spatial homogeneity, or the idea that the expected number of cells in any sample subset is consistent across the sample, called complete
spatial randomness (CSR), $K_0(r)$ has a theoretical value of $\pi r^2$ \citep{baddeley2015spatial}. However, homogeneity is often severely violated in spatial proteomics due to common issues such as tissue degradation over time or incomplete adherence to slides during imaging. When using $\hat{K}_s(r)$ without modification, these inhomogeneities often lead to overestimation of spatial clustering and colocalization and may bias downstream modeling.


\subsection{Scientific motivation and context in ovarian cancer} \label{sec:science}

Ovarian cancer is highly lethal and ranks as the sixth leading cause of cancer-related death among women in the United States. More than 20\% of those diagnosed die within the first year, and only 51\% survive beyond five years \citep{siegel2024cancer}. These facts highlight the urgent need for research that can lead to more personalized care, particularly through a better understanding of the immunological mechanisms underlying the disease.

Our motivating data come from a spatial proteomics study of 128 women with high-grade serous ovarian cancer (HGSOC), in which one tumor sample was collected from each patient at diagnosis, along with overall survival and clinical variables such as age and cancer stage \citep{jordan2020capacity, steinhart2021spatial}. Each sample contains thousands of cells labeled with protein markers identifying tumor cells, immune cells, and key immune cell subtypes including B cells, macrophages, and cytotoxic T cells.

Our scientific focus is on the spatial colocalization of B cells and macrophages. Tumor-associated macrophages (TAMs), the most prevalent immune cell type in HGSOC tumors, are known to promote an inflammatory microenvironment that supports tumor progression. Elevated macrophage infiltration has been linked to poor prognosis in HGSOC \citep{steinhart2021spatial, nowak2020role}. Despite their abundance, the tumor-promoting properties of TAMs remain poorly understood, and few strategies exist for effectively depleting them from the HGSOC microenvironment \citep{lan2013expression}. The role of B cells is even less well characterized. While B cells are traditionally known for antibody production, they also secrete cytokines and engage in paracrine signaling with nearby immune cells, including macrophages. These interactions may be either tumor-promoting or tumor-suppressive, but their functional impact in HGSOC remains unclear \citep{gupta2019b}. Our analysis aims to contribute to this important area of study.

\begin{figure}[h!]
	\centering
	\includegraphics[width=\columnwidth]{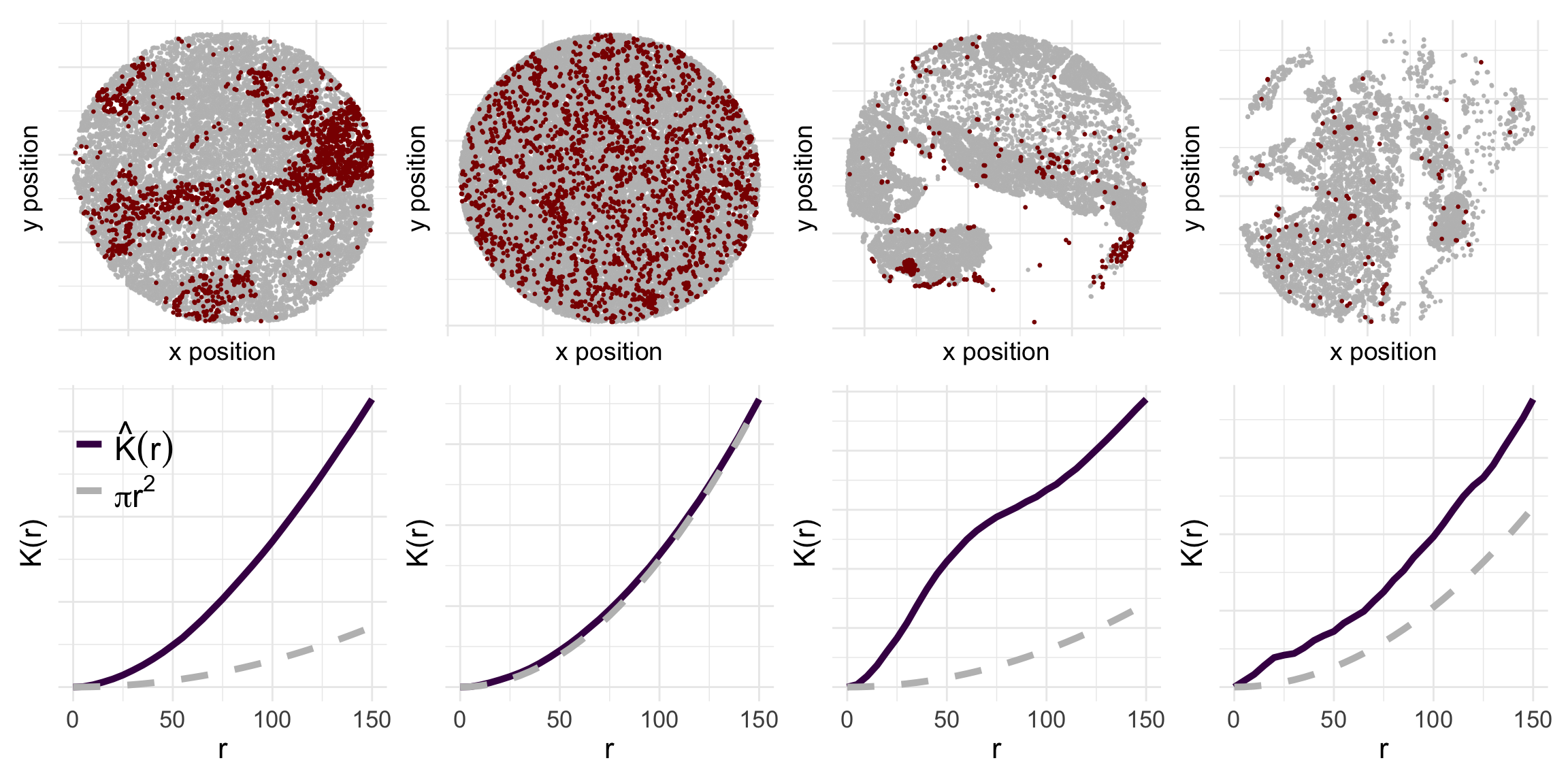}
	\caption{Observed point patterns for four samples from the HGSOC data. Top row shows spatial locations of background cells (gray) and immune cells (red). The bottom row shows the corresponding empirical Ripley's $K$ as a function of radius $r$ (purple) and $K_0(r) = \pi r^2$ (gray). The first and second columns show a representative sample with high and low immune cell clustering, respectively. The third and fourth columns show samples with areas of spatial inhomogeneity due to tearing and degradation of the tissue.}
	\label{fig:patchy_images}
\end{figure}

Representative samples from our dataset are shown in Figure \ref{fig:patchy_images}, which displays observed point patterns (top row) alongside their corresponding $\hat{K}_s(r)$ values (bottom row) for four patient samples. In the point patterns, immune cells are shown in red and all other cells (denoted ``background'' cells) in gray. These panels illustrate the use of Ripley's $K$ to quantify spatial clustering and highlight challenges posed by spatial inhomogeneity due to tissue degradation. In the first column, $\hat{K}_s(r)$ exceeds the theoretical null $K_0(r) = \pi r^2$, indicating spatial clustering of immune cells. The second column shows a sample with similar cell density but more randomly dispersed immune cells, resulting in $\hat{K}_s(r) \approx \pi r^2$.

The patient's sample in the third column of Figure \ref{fig:patchy_images} has areas of spatial inhomogeneity due to tearing and degradation of the tissue. Here, immune cells appear clustered, with $\hat{K}_s(r)$ again exceeding $\pi r^2$ across radii. However, the sample violates the spatial homogeneity assumption necessary for $\pi r^2$ to serve as a valid null, making it unclear whether the elevated $\hat{K}_s(r)$ values reflect true biological clustering or technical artifacts. In the fourth column, the tissue is also degraded, but immune cells are more randomly distributed, suggesting that signal in $\hat{K}_s(r)$ is more likely attributable to inhomogeneity rather than clustering.


\subsection{Our contributions} \label{sec:contribution}

The challenges discussed in Section \ref{sec:science} underscore the need for a method that can adjust for spatial inhomogeneity while preserving meaningful biological signal. In particular, we seek to enable scientifically valid inference about the spatial colocalization of B cells and macrophages and its association with patient survival in HGSOC.

To this end, we propose a novel, computationally efficient $K$-based method. Our method, designed to be robust against sample inhomogeneity and termed \textbf{K} adjustment by \textbf{A}nalytical \textbf{M}oments of the \textbf{P}ermutation distribution (\textit{KAMP}), leverages the first and second moments of the distribution of Ripley's $K$ under all possible permutations of cell labels. \textit{KAMP} serves two main analytical goals in our HGSOC dataset: (1) The first moment provides an empirical estimate of $K_0(r)$ for each sample. By subtracting this estimate from the observed $\hat{K}_s(r)$, we obtain a measure of spatial clustering that can be extracted as a covariate for downstream survival analysis. (2) The second moment (variance) enables the construction of an asymptotically normal test to determine whether $\hat{K}_s(r)$ significantly deviates from $K_0(r)$ for each sample.

As a whole, our proposed \textit{KAMP} method enables accurate spatial analysis of proteomics data both within and across samples while accounting for inhomogeneity. Given the scale of our data, which includes over 1.5 million cells, computational efficiency is crucial. Fortunately, by deriving analytic expressions for these moments and asymptotic properties, our method eliminates the need for actual permutations, making it fast. The codes for implementing the proposed method and reproducing the simulations are available on GitHub at \url{https://github.com/julia-wrobel/KAMP-paper}.

The remainder of this paper is structured as follows: Section \ref{sec:lit_review} presents a review of relevant literature; Section \ref{sec:methods} details our method; Section \ref{sec:simulations} shows simulation results, and in Section \ref{sec:results} we analyze the HGSOC data. We conclude with a discussion in Section \ref{sec:discussion}.


\subsection{Related works} \label{sec:lit_review}

Several spatial summaries can be used to quantify cell-type-specific spatial organization in point-pattern and spatial omics data. For example, the $G$ function summarizes nearest-neighbor distances and is useful for characterizing local nearest-neighbor structure \citep{ripley1988statistical, diggle2013statistical}, while neighborhood enrichment tests, implemented in \texttt{Squidpy}, assess whether cell-type adjacencies are enriched on a pre-specified spatial graph, such as a $k$-nearest-neighbor or radius-based graph \citep{palla2022squidpy}. These approaches address related but distinct scientific questions. We focus on Ripley's $K$ because our scientific target is radius-dependent cumulative clustering and colocalization, for which $K$ has a direct interpretation as the expected number of same-type or cross-type cells within distance $r$, after normalizing for cell density. Moreover, $K$-based summaries have been widely used in spatial proteomics and downstream survival analyses. This makes Ripley's $K$ a natural starting point for developing a scalable empirical-null adjustment for spatial inhomogeneity in existing spatial proteomics workflows.

One approach to addressing inhomogeneity in point pattern data is to use an inhomogeneous adaptation of Ripley's $K$ function, introduced by \citet{baddeley2000non}. This method corrects for inhomogeneity by applying differential weights to the $K$ function at each point, based on the local density of points in the neighborhood of that point. However, because both the weights and the $K$ function are estimated from the same point pattern (represented by the red dots in Figure \ref{fig:patchy_images}, with the gray background cells ignored), this metric may introduce bias \citep{shaw2021globally}.

An alternative method for addressing inhomogeneity is to leverage the background cells in each sample to estimate an empirical null value of $K_0(r)$.
Specifically, an empirical $K_0(r)$ can be estimated by permuting the labels of the immune cells and background cells to generate a null distribution of the $K$ function \citep{wilson2022tumor}. The degree of clustering statistic, $\tilde{K}_s(r)$, is then adjusted by subtracting this empirical $K_0(r)$ from the observed $\hat{K}_s(r)$ rather than using the theoretical value under homogeneity of $K_0(r) = \pi r^2$. While this permutation approach is appealing because it leverages information from surrounding tissue to address the inhomogeneity, its effectiveness in detecting clustering in spatial proteomics data has not been thoroughly evaluated. Furthermore, it becomes computationally inefficient when the number of cells per sample is large. Monte Carlo procedures with a smaller number of randomly chosen permutations can be employed to approximate the permutation distribution, but this approach inherently introduces sampling errors. Our dataset includes 128 samples, each with over 10,000 cells on average. As spatial proteomics becomes more affordable and widely adopted, both the number of samples and cells per sample are expected to increase substantially. Consequently, relying on permutations to estimate the empirical null is becoming increasingly impractical in practice.

This computational issue under the permutation framework is also prevalent in other applications, such as two-sample testing problems for point patterns \citep{hahn2012studentized} and inference for the general linear model in neuroimaging \citep{winkler2014permutation}. To overcome this computational burden, there have been several attempts to approximate the permutation distribution directly in other topics, however, they have some potential drawbacks. A normal approximation with the use of a transformation was studied \citep{heo1998permutation, abdi2007rv}, but it has been noted that the permutation distribution of many statistics is often skewed \citep{mielke1984} and certain transformations consider only a few moments, resulting in suboptimal approximations. A Pearson type III approximation was proposed for the permutation distribution of association in two-by-two tables, analysis of variance statistics in paired genomic data \citep{minas2014distance}, and for kernel association tests in microbiome analysis \citep{zhan2017fast}. However, this approach is heuristic and there is no theoretical foundation. A Davies' method can be applied when the permutation distribution is skewed \citep{davies1980distribution, shinohara2020distance, liu2021kernel}, but it still provides excessive computational burden and is overly conservative \citep{song2022fast}. To our knowledge, this is the first closed-form characterization of the first two permutation-null moments, together with an asymptotic normal approximation, for univariate and bivariate Ripley's \textit{K}.


\section{Methods} \label{sec:methods}

Throughout, we use the term ``spatial inhomogeneity'' to refer to systematic spatial variation in the local cell density across the observation window, equivalently to a non-constant first-order intensity function $\lambda(u)$ of the underlying spatial point process. This includes both biological sources, such as tissue architecture, and technical sources, such as tearing, folding, or partial tissue loss during slide preparation.

Standard applications of Ripley's $K$ rely on the assumption of CSR, where the expected value under the null hypothesis is $\pi r^2$. However, spatial proteomics data frequently exhibit spatial inhomogeneity due to tissue architecture or technical artifacts, violating the homogeneity assumption required for CSR. Consequently, comparing the observed $K$-function against $\pi r^2$ often confounds true biological aggregation with background spatial variation, leading to spurious estimates of clustering. 

To address this, we propose to evaluate ``relative" clustering with respect to the background cell distribution. Let $n$ denote the total number of cells and $m$ denote the number of immune cells in a generic sample. We define our null hypothesis $(H_0)$ as proportional intensity, positing that the spatial intensity of immune cells, $\lambda_{m}(u)$, is proportional to that of the background cells, $\lambda_{n-m}(u)$, such that $\lambda_{m}(u)\propto\lambda_{n-m}(u)$ at any location $u$. This hypothesis implies that immune cells are distributed as a random subset of the overall cell population, following the same spatial structure as the background cells.

Let $\hat{K}_s(r)$ represent the estimated univariate Ripley's $K$ for sample $s$ at radius $r$, and let $n_s$ and $m_s$ denote the total number of cells and the number of immune cells in sample $s$, respectively. $\hat{K}_s(r)$ is the empirical cumulative distribution of the number of pairwise distances between immune cells in the sample that are less than $r$, normalized by the density of cells in the image. Intuitively, smaller pairwise distances between cells are indicative of clustering. Mathematically, $\hat{K}_s(r)$ is given by
\begin{equation}
	\hat{K}_s(r) = \frac{|A|}{m_s(m_s-1)} \sum_{i=1}^{m_s} \sum_{j \neq i}^{m_s}\mathbbm{1}\left(d\left\{c_i, c_j\right\} \leq r\right) e_{i j},
	\label{eq:K}
\end{equation}
where $d\left\{c_i, c_j\right\}$ is the pairwise distance between immune cells $c_i$ and $c_j$, $|A|$ is the tissue area, and $\mathbbm{1}(\cdot)$ is an indicator function \citep{baddeley2005spatstat}. The term $e_{ij}$ is an edge correction to account for bias that occurs when points at the boundary of the tissue region have fewer neighboring points compared to points near the center of the tissue. Several types of edge corrections exist, and their relative merits are discussed elsewhere \citep{ripley1988statistical,baddeley2015spatial}. For our purposes, we assume that these values are symmetric such that $e_{ij} = e_{ji}$. When used to quantify spatial clustering of a single cell type, $\hat{K}_s(r)$ in Equation (\ref{eq:K}) is referred to as ``univariate Ripley's $K$''.

The bivariate form of Ripley's $K$ for quantifying spatial colocalization of two types of immune cells, denoted $\hat{K}_{sc}(r)$, is given by 
\begin{equation}
	\hat{K}_{sc}(r) = \frac{|A|}{m_{s1}m_{s2}} \sum_{i=1}^{m_{s1}} \sum_{j=1}^{m_{s2}}\mathbbm{1}\left(d\left\{c_{1i}, c_{2j}\right\} \leq r\right) e_{i j},
	\label{eq:Ksc}
\end{equation}
where $m_{s1}$ and $m_{s2}$ represent the number of cells of immune cell type 1 and immune cell type 2, respectively, out of a total of $n_s$ cells in sample $s$. $\hat{K}_{sc}$ measures the expected number of immune cells for type 2 found within the distance $r$ from a randomly selected immune cell of type 1 \citep{lagache2013statistical}.


\subsection{Permutation-null moments of Ripley's $K$} \label{subsec:expect}

Under the null hypothesis of proportional intensity, we adopt a label-permutation approach that randomly assigns immune/background labels to the observed pooled locations while keeping the locations fixed. This procedure constructs an empirical null benchmark that inherently accounts for the sample's underlying spatial inhomogeneity.

For univariate analysis, each sample contains $n_s$ total cells, with $m_s$ immune cells and $n_s-m_s$ ``background'' cells. Our goal is to obtain the expectation and variance of the distribution of $\hat{K}_s(r)$ value obtained when permuting the labels of immune and background cells. For bivariate analysis (measuring spatial colocalization of two immune cell types), suppose each image contains $n_s$ total cells, with $m_{s1}$ immune cells for type 1 and $m_{s2}$ immune cells for type 2, and $n_s-(m_{s1}+m_{s2})$ background cells.

To assist in this derivation, Equation (\ref{eq:K}) can be expressed in a matrix form as follows:
\begin{equation}
	\hat{K}_s(r) = \frac{|A|}{m_{s}(m_{s}-1)}\mathbf{x}^T\mathbf{W}(r)\mathbf{x},  
	\label{eq:Kmat}
\end{equation}
where $\mathbf{x}$ is an $n_s\times 1$ vector of 1's and 0's indicating whether each cell is an immune cell, and $\mathbf{W}(r)$ is an $n_s\times n_s$ matrix whose $(i,j)$-th entry is given by $\mathbbm{1}\left(d\left\{c_i, c_j\right\} \leq r\right) e_{i j}$. 
For $n_s$ total cells, there will be $n_s!$ possible permutations of the cell labels.

Let $\mathbf{P}$ represent an $n_s\times n_s$ permutation matrix, e.g. a square, binary matrix with one entry of $1$ in each row and each column, and $0$'s elsewhere. For each possible permutation of the cell labels, $p$, there exists a specific permutation matrix $\mathbf{P}_p; p\in (1,\ldots,n_s!)$ such that $\mathbf{P}_p\times \mathbf{x}$ provides a vector of permuted labels. For example, $(|A|m_s^{-1}(m_s-1)^{-1})\mathbf{x}^T\mathbf{P}_p^{T}\mathbf{W}(r)\mathbf{P}_p\mathbf{x}$ is equivalent to Equation (\ref{eq:Kmat}) when $\mathbf{P}_p = I$ for the $n_s\times n_s$ identity matrix $I$.

Similarly, Equation (\ref{eq:Ksc}) can be expressed in the matrix form as follows:
\begin{equation}
	\hat{K}_{sc}(r) = \frac{|A|}{m_{s1}m_{s2}}\mathbf{x}_{1}^T\mathbf{W}(r)\mathbf{x}_{2} = \frac{|A|}{m_{s1}m_{s2}}\mathbf{x}_{1}^T\mathbf{P}_p^T\mathbf{W}(r)\mathbf{P}_p\mathbf{x}_{2},  
	\label{eq:Kmatsc}
\end{equation}
where $\mathbf{x}_{1}$ and $\mathbf{x}_{2}$ are $n_s\times 1$ vectors of 1's and 0's indicating whether each cell is an immune cell for type 1 and for type 2, respectively, and $\mathbf{P}_p = I_{n_s\times n_s}$.

To obtain the analytic formulas for the expectation and variance of $K$ under the permutation null distribution, we rewrite Equation (\ref{eq:K}) in the following way. For $n_s$ total cells, let $g_{i}=1$ if a cell $c_{i}$ is an immune cell and 0 otherwise. Then, 
\begin{equation} 
	\hat{K}_s(r) = \frac{|A|}{m_{s}(m_{s}-1)}\sum_{i=1}^{n_{s}}\sum_{j\ne i}^{n_{s}}W_{ij}(r)\mathbbm{1}\left(g_{i}=g_{j}=1\right),
\end{equation}
where $W_{ij}(r)$ is $(i,j)$-th  entry of $\mathbf{W}(r)$, that is, $W_{ij}(r) = \mathbbm{1}\left(d\left\{c_i, c_j\right\} \leq r\right)e_{ij}$. Similarly, Equation (\ref{eq:Ksc}) can be rewritten as follows:
\begin{equation} 
	\hat{K}_{sc}(r) = \frac{|A|}{m_{s1}m_{s2}}\sum_{i=1}^{n_{s}}\sum_{j\ne i}^{n_{s}}W_{ij}(r)\mathbbm{1}(g'_{i}=1, g'_{j}=2),
\end{equation}
where $g'_{i}=1$ if a cell $c_{i}$ is an immune cell for type 1, $g'_{i}=2$ if a cell $c_{i}$ is an immune cell for type 2, and 0 otherwise.

To make the permutation distribution explicit, let $P$ denote a uniformly sampled permutation matrix from all possible permutations of the $n_s$ cell labels. We define $K_s(r)$ and $K_{sc}(r)$ as the random variables obtained by evaluating the univariate and bivariate Ripley's $K$ statistics after applying this random label permutation:
	\begin{align*}
		K_s(r) &= \frac{|A|}{m_s(m_s-1)} x^\top P^\top W(r)P x, \\
		K_{sc}(r) &= \frac{|A|}{m_{s1}m_{s2}} x_1^\top P^\top W(r)P x_2.
	\end{align*}
	The observed statistics $\hat K_s(r)$ and $\hat K_{sc}(r)$ correspond to the realized labeling, equivalently the identity permutation $P=I$. Let $\ep$ and $\var$ be the expectation and variance under the permutation distribution. The analytic expressions for the expectation and variance of $K$ by random permutations are provided in the following theorem.
\begin{theorem} \label{thm:expect}
	Under the permutation null distribution, we have
	{\small
	\begin{align*}
		\ep\left(K_{s}(r)\right) &= \ep\left(K_{sc}(r)\right) = \frac{|A|}{n_{s}(n_{s}-1)}R_{0}, \\
		\var\left(K_{s}(r)\right) &= \frac{|A|^2}{m_{s}^2(m_{s}-1)^2}\left\{2R_{1}f_{1}(m_{s}) + 4R_{2}f_{2}(m_{s}) + R_{3}f_{3}(m_{s})\right\} - \ep^2\left(K_{s}(r)\right), \\
		\var\left(K_{sc}(r)\right) &= \frac{|A|^2}{m_{s1}^2m_{s2}^2}\left\{R_{1}h_{1}(m_{s1},m_{s2}) + R_{2}h_{2}(m_{s1},m_{s2}) + R_{3}h_{3}(m_{s1},m_{s2})\right\} - \ep^2\left(K_{sc}(r)\right),
	\end{align*}}
	where
	{\small
	\begin{align*}
		f_1(x) &= \frac{x(x-1)}{n_{s}(n_{s}-1)}, \ \ f_2(x)  = \frac{x(x-1)(x-2)}{n_{s}(n_{s}-1)(n_{s}-2)}, \ \ f_3(x) = \frac{x(x-1)(x-2)(x-3)}{n_{s}(n_{s}-1)(n_{s}-2)(n_{s}-3)}, \\
		h_1(x,y) &= \frac{xy}{n_{s}(n_{s}-1)}, \ \ h_2(x,y)  = \frac{xy(x+y-2)}{n_{s}(n_{s}-1)(n_{s}-2)}, \ \ h_3(x,y) = \frac{xy(x-1)(y-1)}{n_{s}(n_{s}-1)(n_{s}-2)(n_{s}-3)}, \\[5mm]
		R_{0} &= \sum_{i=1}^{n_{s}}\sum_{j\ne i}^{n_{s}}W_{ij}(r), \ \ R_{1} = \sum_{i=1}^{n_{s}}\sum_{j\ne i}^{n_{s}}W_{ij}^2(r), \ \ R_{2} = \sum_{i=1}^{n_{s}}\sum_{j\ne i}^{n_{s}}\sum_{u\ne i\ne j}^{n_{s}}W_{ij}(r)W_{iu}(r), \\ 
		R_{3} &= \sum_{i=1}^{n_{s}}\sum_{j\ne i}^{n_{s}}\sum_{u\ne i\ne j}^{n_{s}}\sum_{v\ne u\ne i\ne j}^{n_{s}}W_{ij}(r)W_{uv}(r).
	\end{align*}}
\end{theorem}
Based on Theorem \ref{thm:expect}, the new degree of clustering can be defined as $\tilde{K}_{s}(r) = \hat{K}_{s}(r) - \ep\left(K_{s}(r)\right)$ or $\tilde{K}_{sc}(r) = \hat{K}_{sc}(r) - \ep\left(K_{sc}(r)\right)$.

It follows from the result in Theorem \ref{thm:expect} that the expected value of both $K_s(r)$ and $K_{sc}(r)$ under all permutations of the cell labels is given by
$\ep\left(K_{s}(r)\right) = \ep\left(K_{sc}(r)\right) = |A|\{n_{s}(n_s-1)\}^{-1}\sum_{i=1}^{n_{s}}\sum_{j\ne i}^{n_{s}}W_{ij}(r) = |A|\{n_{s}(n_s-1)\}^{-1} \sum_{i=1}^{n_s} \sum_{i \neq j}^{n_s}\mathbbm{1}\left(d\left\{c_i, c_j\right\} \leq r\right) e_{i j}$. This result is noteworthy because it shows that the expectation of both $K_s(r)$ and $K_{sc}(r)$ under all permutations is equivalent to the univariate Ripley's $K$ computed using all cells in the sample. Under the permutation null, the cell locations are fixed and only the labels are randomized. Therefore, for any ordered pair of distinct locations $(i,j)$, the probability that the pair is selected as an immune--immune pair in the univariate setting is $m_s(m_s-1)\{n_s(n_s-1)\}^{-1}$ and the probability that it is selected as a type 1--type 2 pair in the bivariate setting is $m_{s1}m_{s2}\{n_s(n_s-1)\}^{-1}$. These probabilities cancel with the corresponding normalizing constants in the definitions of $K_s(r)$ and $K_{sc}(r)$, leaving the common contribution $|A|\sum_{i=1}^{n_s}\sum_{j\ne i} W_{ij}(r)\{n_s(n_s-1)\}^{-1}$. Thus, averaging over all label permutations removes cell-type-specific spatial information and leaves the spatial structure of the pooled set of observed cells. 

As a practical consequence, existing software for Ripley's $K$ can be directly used to compute $\ep\left(K_{s}(r)\right)$ and $\ep\left(K_{sc}(r)\right)$. Theorem \ref{thm:expect} can be proven through combinatorial analysis. However, calculating the variance of $K$ necessitates a more meticulous examination of different combinations. The complete proof of the theorem is provided in Supplement A. 


\subsection{Within-sample inference} \label{subsec:inf}

Given sample $s$ and radius $r$, we conduct one-sided tests for positive excess spatial clustering or colocalization relative to the permutation-null benchmark. For the univariate statistic, the null hypothesis is the proportional-intensity, or random-labeling, null under which cell labels are exchangeable conditional on the observed cell locations and cell counts. The one-sided alternative is positive excess clustering, meaning that $\hat{K}_{s}(r) > \ep(K_{s}(r))$. The bivariate colocalization test is defined analogously, with the one-sided alternative $\hat{K}_{sc}(r) > \ep(K_{sc}(r))$. Based on the results from Section \ref{subsec:expect}, we use the following test statistic to quantify immune cell clustering in the sample:
\begin{equation}
	Z_{K_{s}}(r) = \frac{\hat{K}_{s}(r) - \ep\left(K_{s}(r)\right)}{\sqrt{\var\left(K_{s}(r)\right)}}.
	\label{eq:Z}
\end{equation}
To quantify immune cell colocalization, we substitute $\hat{K}_{s}(r)$, $\ep\left(K_{s}(r)\right)$, and $\var\left(K_{s}(r)\right)$ with $\hat{K}_{sc}(r)$, $\ep\left(K_{sc}(r)\right)$, and $\var\left(K_{sc}(r)\right)$, respectively, in Equation (\ref{eq:Z}) to obtain $Z_{K_{sc}}(r)$.
Large positive values of $Z_{K_{s}}(r)$ and $Z_{K_{sc}}(r)$ provide evidence against the null hypotheses.

We next show that the asymptotic distributions of $Z_{K_{s}}(r)$ and $Z_{K_{sc}}(r)$ are standard normal under the permutation null distribution.  In the following, we write $a_N \asymp b_N$ when $a_N$ has the same order as $b_N$ and $a_N = o(b_N)$ when $a_N$ is dominated by $b_N$ asymptotically, i.e., $\lim_{N\rightarrow\infty}(a_{N}/b_{N}) = 0$. Let $\tilde{W}_{ij}(r) = \left(W_{ij}(r) - \bar{W}(r)\right)\mathbbm{1}_{i\ne j}$, $\bar{W}(r) = n_{s}^{-1}(n_{s}-1)^{-1}\sum_{i=1}^{n_{s}}\sum_{j\ne i}^{n_{s}}W_{ij}(r)$, and $\tilde{W}_{i\cdot}(r) = \sum_{j=1,j\ne i}^{n_{s}}\tilde{W}_{ij}(r)$ for $i=1,\ldots,n_{s}$. We work under the following two conditions:
\begin{condition} \label{condition1}
	$\sum_{i=1}^{n_{s}}|\tilde{W}_{i\cdot}(r)|^h = o(\{\sum_{i=1}^{n_{s}}\tilde{W}_{i\cdot}^2(r)\}^{h/2})$ for all integers $h>2$.
\end{condition}
\begin{condition} \label{condition2}
	$\sum_{i=1}^{n_{s}}\sum_{j=1,j\ne i}^{n_{s}}\tilde{W}_{ij}^2(r) = o(\sum_{i=1}^{n_{s}}\tilde{W}_{i\cdot}^2(r))$.
\end{condition}
\begin{theorem} \label{thm:asymp}
	Under the permutation null distribution, as $n_{s}\rightarrow\infty$, $m_{s}/n_{s} \rightarrow p_{s} \in (0,1)$, $m_{s1}/n_{s} \rightarrow p_{s1} \in (0,1)$, $m_{s2}/n_{s} \rightarrow p_{s2} \in (0,1)$, $Z_{K_{s}}(r)$ and $Z_{K_{sc}}(r)$ converge in distribution to the standard normal distribution under Conditions \ref{condition1} and \ref{condition2}.
\end{theorem}

\begin{remark}
	Condition \ref{condition1} can be satisfied when $|\tilde{W}_{i\cdot}(r)| = O(n_{s}^{\delta})$ for a constant $\delta$, $\forall i$, and Condition \ref{condition2} would further be satisfied if we also have $\tilde{W}_{ij}(r) = O(n_{s}^{\kappa})$ for a constant $\kappa < \delta - 0.5$, $\forall i,j$. Noting that Conditions \ref{condition1} and \ref{condition2} are satisfied when $\tilde{W}_{ij}(r)$'s are of constant order. Moreover, some of the values $W_{ij}(r)$ would be zero, depending on the choice of $r$. Since $\tilde{W}_{ij}(r) = W_{ij}(r) -\bar{W}(r)$, unless there are exceptional circumstances of significant outliers, Conditions \ref{condition1} and \ref{condition2} are generally satisfied.
\end{remark}

\begin{remark}
	Scientifically, Condition \ref{condition1} can be viewed as a no-dominating-cell condition: the centered neighborhood contribution should not be driven by a single cell or a small number of cells. This means that the \textit{KAMP} statistic should reflect spatial organization accumulated across many cells, rather than being driven by a few cells in an unusually dense region, a tissue artifact, or boundary cells with extreme edge corrections. Condition \ref{condition2} requires that cell-level neighborhood variation dominates individual pairwise contributions, so that the statistic is driven by local neighborhood structure rather than by a small number of close-range cell pairs. These conditions are most plausible in spatial proteomics settings with many cells per sample and radii chosen to capture local neighborhoods within the tissue region.
\end{remark}

The full proof of Theorem \ref{thm:asymp} is provided in Supplement B. These results enable us to perform explicit hypothesis tests for spatial clustering and colocalization that yield approximate $p$-values. Furthermore, for a fixed immune cell abundance $p$, we expect the statistical properties of these tests to improve as the total number of cells $n_s$ increases.

\begin{figure}[h!]
	\centering
	\includegraphics[width=\columnwidth]{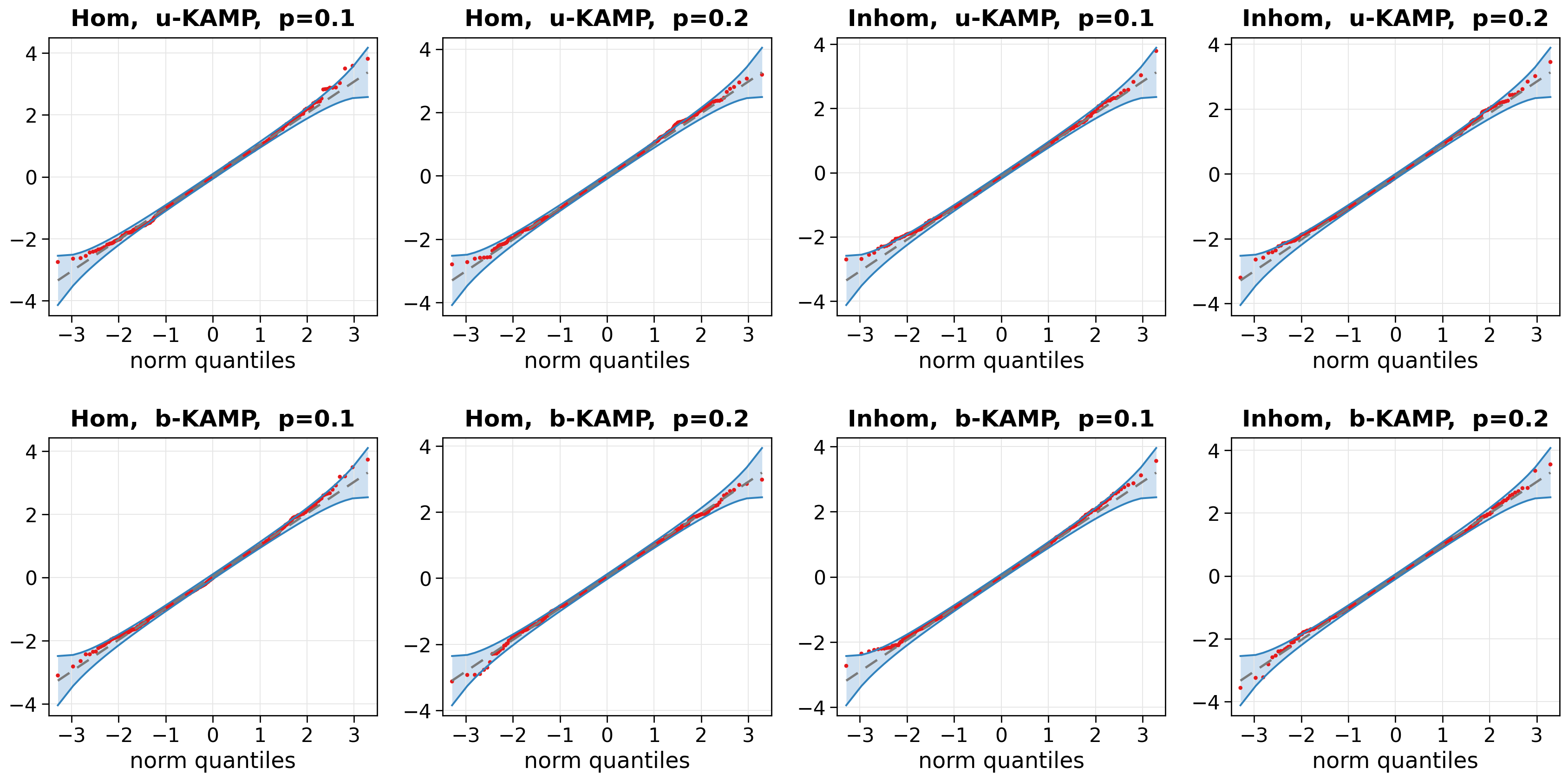}
	\caption{Normal quantile-quantile plots of univariate KAMP statistics (u-KAMP) and bivarate KAMP statistics (b-KAMP) under different situations where $\lambda_{n}=2000$.}
	\label{fig:qq}
\end{figure}

Figure \ref{fig:qq} shows normal quantile-quantile plots for the univariate and bivariate \textit{KAMP} statistics from 1,000 permutations under two abundance values and homogeneous and inhomogeneous null conditions described in Table 1, with $\lambda_{n}=2000$. ``Hom” and ``Inhom” represent the conditions where cells were generated under homogeneity or inhomogeneity, respectively, described in Section \ref{subsec:design}. Figure \ref{fig:qq} indicates that the permutation distributions can be well approximated by the standard normal distribution under both homogeneous and inhomogeneous null conditions.


\subsection{Computational improvements and patient-level analysis}

As $n_s$, the total number of cells in sample $s$, increases, computation time also rises because estimating $\ep\left(K_{s}(r)\right)$ and $\var\left(K_{s}(r)\right)$ requires evaluating pairwise relationships between cells. To improve computational efficiency, our implementation avoids constructing dense $n_s \times n_s$ distance and edge-correction matrices. Instead, neighboring cell pairs within radius $r$ are identified using a KD-tree range query, and the edge-correction weights and quantities required to compute the analytic permutation moments are evaluated only for these nearby pairs using sparse vectorized operations. This substantially reduces both memory usage and computation time.

For very large spatial proteomics datasets, computation can be reduced even further through independent random thinning of the points in sample $s$. When a point process is randomly thinned by independently selecting points for removal, the resulting point process retains the same distributional properties as the original \citep{baddeley2015spatial}. Therefore, thinning the cells before computing $\ep\left(K_s(r)\right)$ and $\var\left(K_s(r)\right)$ provides accurate approximations while substantially reducing computation time. We refer to this computationally efficient approximation as \textit{KAMP lite} throughout the remainder of the paper.

After quantifying spatial clustering or colocalization in each sample, the next objective is to analyze these metrics across samples, typically by assessing their association with patient-level outcomes. In the HGSOC application below, we use the bivariate summary $\tilde{K}_{sc}(r)$ for B cell–macrophage colocalization as a covariate in Cox proportional hazards models to quantify its association with overall survival. Analogous patient-level models could be used for univariate immune-cell clustering or other cell-type pairs. Although we focus on a survival outcome and evaluate $\tilde{K}_s(r)$ at each radius over a grid, other types of outcomes could be analyzed, or $\tilde{K}_s(r)$ over a range of radii could be treated as a functional covariate as in \cite{vu2022spf}.


\section{Simulation studies} \label{sec:simulations}

In this section, we evaluate the performance of our method and competing approaches across a range of settings using simulations designed to mirror the structure of our motivating data. Data are generated from spatial point processes under varying homogeneity assumptions. We consider both univariate and bivariate versions of our method: in the univariate setting, the goal is to identify spatial clustering of a single cell type, while in the bivariate case, we aim to detect colocalization between two distinct cell types. For the univariate scenarios, we simulate point processes involving two types of points: background cells and immune cells. For the bivariate scenarios, we simulate three types of points: background cells, immune cell type 1, and immune cell type 2. In describing the simulation design and comparative methods, we primarily focus on the univariate case; unless otherwise specified, the bivariate case follows a similar approach.


\subsection{Simulation design} \label{subsec:design}

Let $\lambda_n$ and $\lambda_m$ represent the intensities of total cells and immune cells, respectively, within the sample. The expected proportion, or \textit{abundance}, of immune cells is represented by $p = \lambda_{m}/\lambda_{n}$. To evaluate various scenarios, we simulate point process data under the four homogeneity conditions outlined in Table \ref{fig:simConditions}, which reflect the patterns observed in the HGSOC data in Figure \ref{fig:patchy_images}. The first two conditions represent data generated under the null hypothesis of no immune cell clustering, while the last two are generated under the alternative hypothesis of immune cell clustering.

\begin{table}[h!]
	\centering
	\caption{The four spatial organization conditions under which simulated data are generated}
	\includegraphics[width=\columnwidth]{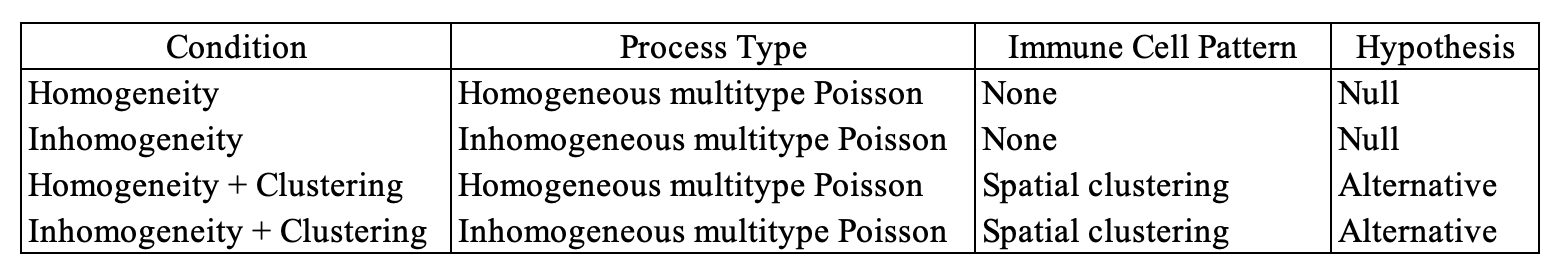}
	\label{fig:simConditions}
\end{table}

A multitype Poisson process can be viewed as a collection of independent Poisson point processes, one for each cell type, defined on the same observation window. Each cell type has its own intensity function. In our univariate null simulations, the two types correspond to immune and background cells; in the bivariate null simulations, they correspond to background cells and two immune-cell types. Null scenarios are simulated from multitype Poisson processes using \texttt{spatstat} \citep{baddeley2015spatial}. Under the null homogeneous condition, the intensities of background and immune cells are given by $\lambda_n - \lambda_m$ and $\lambda_m$, respectively. Under the null inhomogeneous condition, background and immune cells are first generated from homogeneous multitype Poisson processes with intensities $\lambda_n-\lambda_m$ and $\lambda_m$, respectively. Random ``holes'' are then introduced into each sample to mimic the tissue tearing and degradation artifacts commonly observed in spatial proteomics data. This procedure induces spatial inhomogeneity while preserving the null hypothesis of no immune cell clustering.

For the alternative scenarios, we first generate a single-type Poisson spatial point pattern with intensity $\lambda_n$. Given a specified immune cell abundance $p$, immune cell labels are then assigned with probability $p$ to cells within 25 randomly located spatial clusters, each with an average radius of 1.25. This induces spatial clustering of immune cells while maintaining the abundance $p$ and intensity of immune cells $\lambda_m$. For the alternative inhomogeneous condition, random ``holes'' are introduced into each sample to mimic the tissue tearing and degradation artifacts commonly observed in spatial proteomic samples. This process was implemented using the \texttt{GenerateHoles()} function from the \texttt{scSpatialSim} R package \citep{soupir2025scspatialsim}, a recently developed tool for simulating realistic single-cell spatial molecular data. The function generates a kernel-based probability surface that targets a specified proportion of the spatial domain to be designated as missing. Kernel centers are sampled randomly, and regions near each center are assigned a higher probability of being designated a hole. Cells located in these regions are removed, producing localized gaps in the tissue that resemble degraded or torn sections observed in real data. 

For each of the four conditions in Table \ref{fig:simConditions}, we evaluate performance of our method as a function of total cell intensity $\lambda_n$ and immune cell abundance $p$. We simulate datasets for each combination of the four homogeneity conditions, intensities $\lambda_n \in (1000, 2000, 5000, 10000)$, and abundances $p \in (0.01, 0.1, 0.2)$, resulting in 48 distinct scenarios. To evaluate the computational performance and the efficacy of each method in capturing the degree of clustering, 50 datasets are generated for each scenario. To assess Type I error and power, we simulate 1000 additional datasets for each scenario.

To further evaluate the robustness of \textit{KAMP}, we also consider an additional simulation setting in which both immune and background cells exhibit spatial clustering. Cell labels were generated using Gaussian kernel mixtures implemented in the \texttt{scSpatialSim} package, producing either independent clustering of the two cell populations or shared (dependent) clustering in which both cell types occupy the same spatial regions. These simulations assess whether \textit{KAMP} can distinguish immune-specific clustering from clustering explained by the underlying tissue architecture. Results are presented in Supplement E, where we show that \textit{KAMP} performs well when immune clustering is spatially distinct from background clustering but, as expected, loses power when the two clustering patterns are largely shared.


\subsection{Comparison with competing approaches}

We evaluate our proposed methods in terms of computational efficiency and ability to estimate the degree of clustering, denoted $\tilde{K}_s(r)$, across homogeneity conditions. We compare our methods, \textit{KAMP} and \textit{KAMP lite}, with three competing approaches, which are referred to as \textit{K}, \textit{Kinhom}, and \textit{perm}.
\textit{K} represents the degree of clustering under theoretical spatial homogeneity, where $K_0(r) = \pi r^2$. \textit{Kinhom} is the inhomogeneous variant of $\tilde{K}_s(r)$ introduced by \cite{baddeley2000non}, also with $K_0(r) = \pi r^2$. Unlike our method, \textit{Kinhom} does not utilize background cells to infer additional information about inhomogeneity. The \textit{perm} approach involves the permutation of cell labels, followed by estimation of an empirical $K_0(r)$ by calculating $\hat{K}_s(r)$ for each permutation then averaging across all permutations. In all simulation scenarios, we employed 1000 permutations for the \textit{perm} method. For the \textit{KAMP} method, $K_0(r) = \ep\left(K_{s}(r)\right)$. \textit{KAMP lite} introduces random thinning (at probabilities 0.25, 0.5, 0.75, or 0.9) and uses the remaining cells to approximate $\ep\left(K_{s}(r)\right)$.

We also assess power and Type I error under the null and alternative conditions defined in Table \ref{fig:simConditions}, using $\alpha = 0.05$. For \textit{KAMP} and \textit{KAMP lite}, $p$-values are computed using the inference procedure in Section \ref{subsec:inf}. For \textit{perm}, $p$-values are obtained by comparing the observed $\hat{K}_s(r)$ to its empirical null distribution based on permutations.


\subsection{Simulation results}

\begin{figure}[h!]
	\centering
	\includegraphics[width=\columnwidth]{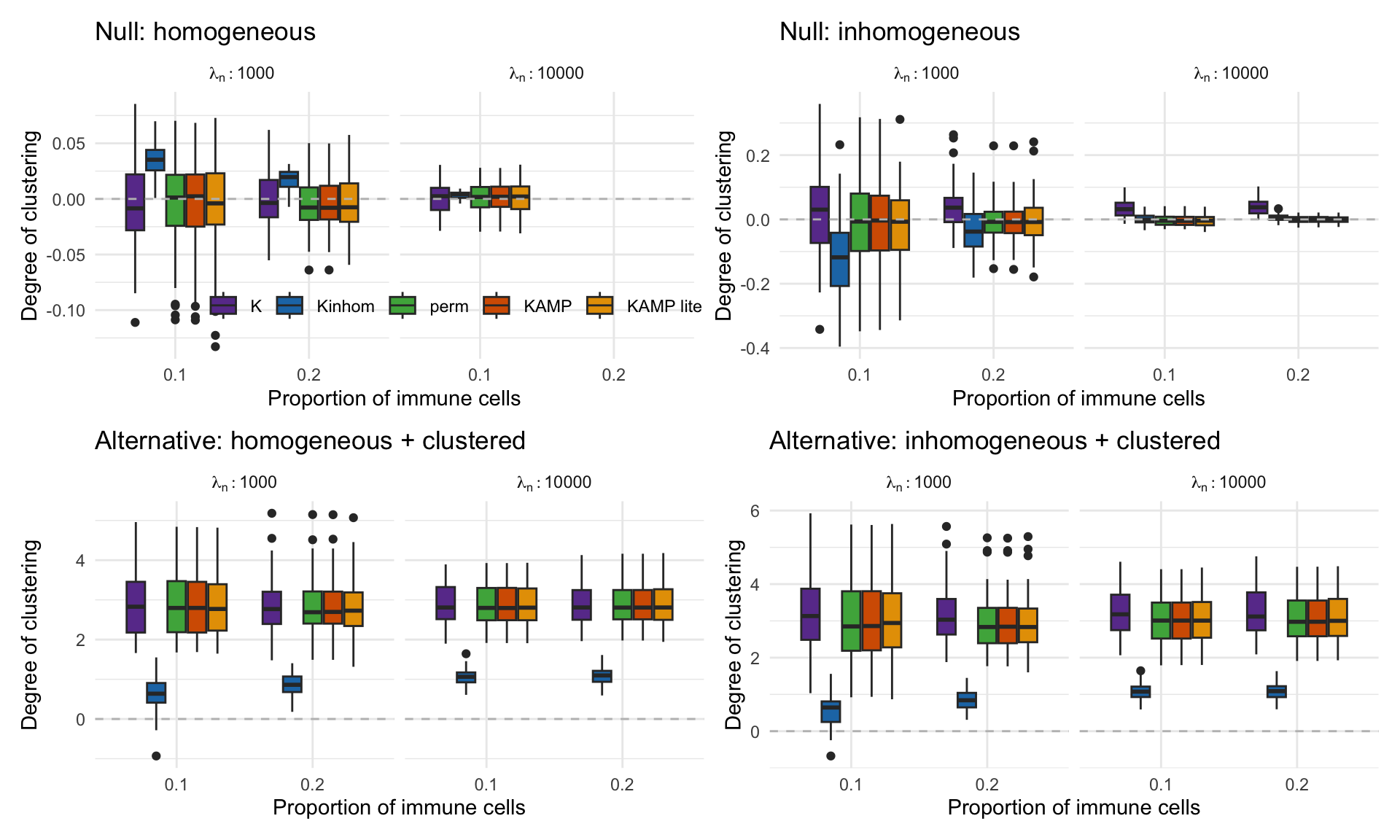}
	\caption{Distribution of $\tilde{K}(r)$ values for \textit{K}, \textit{Kinhom}, \textit{perm}, \textit{KAMP}, and \textit{KAMP lite} across varying immune cell proportions and expected total cell counts $\lambda_n$ under four different spatial homogeneity conditions. The top row shows the null scenarios under homogeneous (left) and inhomogeneous (right) conditions, while the bottom row shows the corresponding alternative scenarios with immune cell clustering. The dotted horizontal line at zero represents no spatial clustering, whereas $\tilde{K}(r) > 0$ indicates positive spatial clustering.}
	\label{fig:sim_results}
\end{figure}

Figure \ref{fig:sim_results} summarizes degree of clustering $\tilde{K}(r)$ results across simulated datasets with varying immune cell abundances $p \in \{0.1, 0.2\}$ and total cell intensities $\lambda_n \in \{1000, 10000\}$. Each panel represents a different spatial homogeneity condition from Table \ref{fig:simConditions}. Across simulation settings and homogeneity conditions, we compare $\tilde{K}(r)$ values for the \textit{K}, \textit{Kinhom}, \textit{perm}, \textit{KAMP}, and \textit{KAMP lite} methods. The dotted horizontal line through zero represents no spatial clustering, whereas $\tilde{K}(r) > 0$ represents clustering. The top two panels represent null clustering conditions, and well-performing methods should have $\tilde{K}(r)$ centered around zero; in contrast, the bottom two panels are conditions with clustering. Unless otherwise specified, \textit{KAMP lite} results are presented using a thinning probability of 0.5.

Across all conditions and methods, the variance of $\tilde{K}(r)$ decreases as both $p$ and $\lambda_n$ increase. In the homogeneous no clustering condition (top left panel), all methods perform well except for \textit{Kinhom}, which systematically overestimates $\tilde{K}(r)$ when $\lambda_n = 1000$ or $p = 0.1$. In the inhomogeneous no clustering condition (top right panel), the \textit{perm}, \textit{KAMP}, and \textit{KAMP lite} methods display distributions of $\tilde{K}(r)$ centered around zero, as expected in a null condition. However, the \textit{K} method's distributions are much higher than zero, illustrating why our method is necessary for obtaining valid estimates of biological clustering in our motivating dataset, since \textit{K} registers all inhomogeneity as clustering. In the homogeneous and clustered condition (bottom left), all methods perform well again, with the exception of \textit{Kinhom}, which underestimates $\tilde{K}(r)$. Finally, in the inhomogeneous and clustered condition (bottom right) \textit{perm}, \textit{KAMP}, and \textit{KAMP lite} perform similarly, whereas \textit{K} shows elevated values across all $\lambda_n$ and $p$ levels, because it is erroneously interpreting inhomogeneity caused by the simulated holes in the data as additional clustering.

\begin{figure}[h!]
	\centering
	\includegraphics[width=\columnwidth]{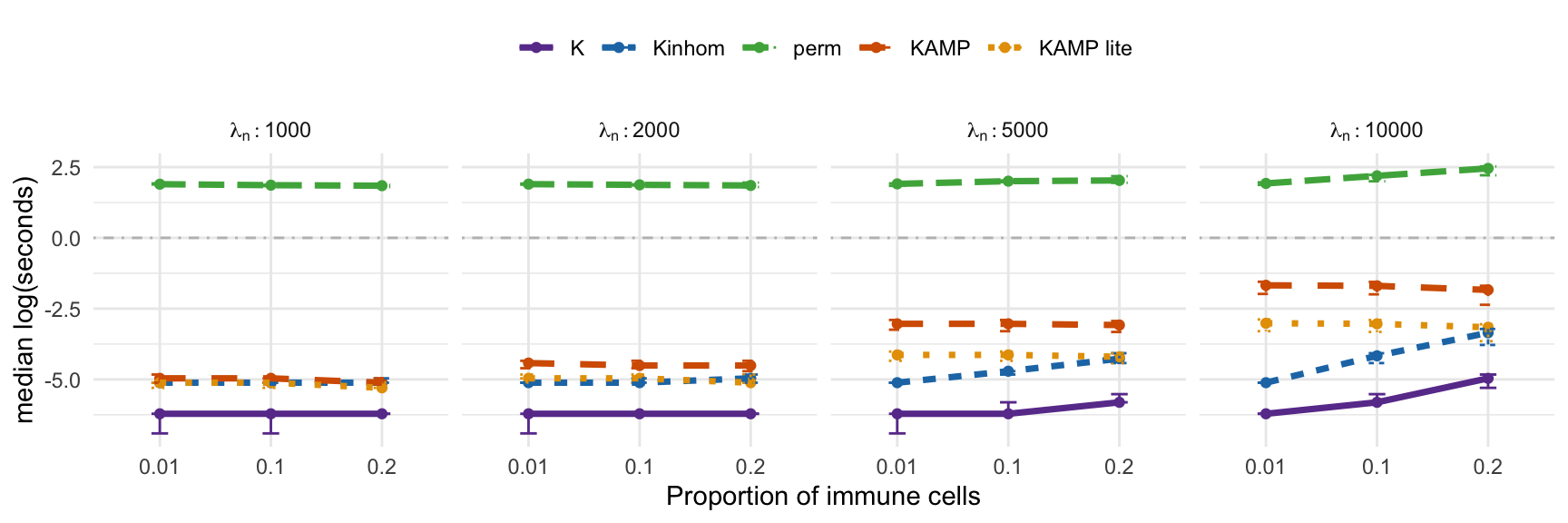}
	\caption{Median computation times (log seconds), with vertical error bars showing the corresponding interquartile range across 50 simulated datasets, for $\tilde{K}(r)$ in the homogeneous, no-clustering condition across varying values of expected number of total cells ($\lambda_n$) and proportion of immune cells ($p$). The dotted line at zero represents a time of one second. These computation times correspond to estimation of the clustering statistic and do not include calculation of the permutation variance and corresponding $p$-value.}
	\label{fig:sim_resultsComp}
\end{figure}

Figure \ref{fig:sim_resultsComp} presents the median computation times in log (seconds) for estimating $\tilde{K}(r)$ in the homogeneous, no clustering condition. These computation times correspond to estimation of the clustering statistic and do not include calculation of the permutation variance and corresponding $p$-value. Across all simulation scenarios, the median computation time for \textit{KAMP} is less than one second. As expected, \textit{KAMP lite} is faster than \textit{KAMP}, as \textit{KAMP lite} effectively reduces $\lambda_n$ by half. \textit{KAMP} and \textit{KAMP lite} consistently outperform \textit{perm} across $\lambda_n$ and $p$ values. However, \textit{KAMP} scales with $\lambda_n$ and \textit{perm} scales with $p$, so for very large $\lambda_n$ and small $p$ additional computational considerations are needed to ensure \textit{KAMP} remains efficient.

\begin{figure}[h!]
	\centering
	\includegraphics[width=\columnwidth]{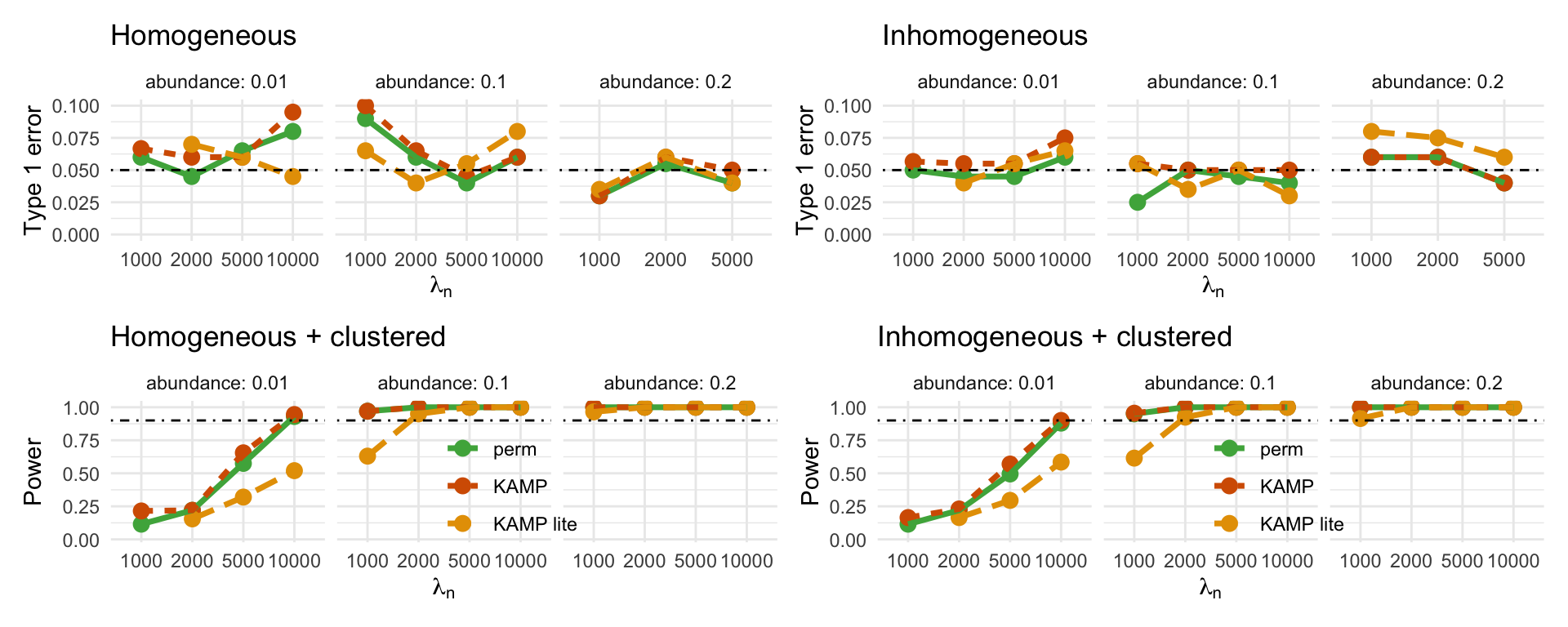}
	\caption{Type I error and power across different values of $p$ and $\lambda_n$ for the \textit{perm}, \textit{KAMP}, and \textit{KAMP lite} methods. The dotted horizontal lines are at 0.05 for the upper panels and 0.9 for the lower panels.}
	\label{fig:sim_resultsPower}
\end{figure}

Figure \ref{fig:sim_resultsPower} displays the Type I error and power across different values of $p$ and $\lambda_n$ for the \textit{perm}, \textit{KAMP}, and \textit{KAMP lite} methods. Across all $\lambda_n$ values and for $p \in \{0.1, 0.2\}$, the \textit{KAMP} method demonstrates strong performance in terms of both power and Type I error, with the \textit{perm} method showing comparable results. In the low abundance scenarios ($p = 0.01$), a larger $\lambda_n$ is required to achieve adequate power, which is expected given that at low $p$ and $\lambda_n$, there may be insufficient immune cells to reliably estimate $\tilde{K}(r)$. The \textit{KAMP lite} method has lower performance compared to \textit{KAMP} in both power and Type I error, although its performance improves as sample size increases. Similar trends are observed in the homogeneous and inhomogeneous conditions.

Additional simulation results assessing the performance of the bivariate \textit{KAMP} method for quantifying spatial colocalization between two distinct cell types are presented in Supplement C. We observe that these results are consistent with the spatial clustering metrics of a single cell type shown in this section.

Further, Supplement D presents simulation results assessing the performance of \textit{KAMP lite} across thinning probabilities in $\{0, 0.25, 0.5, 0.75, 0.9\}$, including univariate and bivariate simulations at low cell abundances. These simulations demonstrate that the degree of clustering $\tilde{K}_s(r)$ remains stable across thinning levels. However, more aggressive thinning increases the variance of the permutation distribution and can reduce power for within-sample hypothesis testing, particularly when the cell types of interest are rare.


\section{Analysis of ovarian cancer spatial proteomics data} \label{sec:results}

We begin by describing data cleaning for the ovarian cancer spatial proteomics data. Next, we conduct within-sample analyses using univariate clustering of immune cells and bivariate colocalization of B cells and macrophages, applying methods \textit{K}, \textit{KAMP}, \textit{KAMP lite}, and \textit{perm}. In the within-sample setting, we focus on univariate clustering of all immune cells to illustrate computational efficiency and compare the distribution of $\tilde{K}_s(r)$ across methods. This approach is computationally demanding due to the large number of immune cells.

We then conduct an across-sample analysis that relates spatial summaries from each image to patient overall survival. While the within-sample analysis emphasizes computational and statistical performance, the across-sample analysis demonstrates the scientific utility of our method. In particular, we introduce a previously unexplored analysis using the bivariate form of \textit{KAMP} to assess how colocalization of B cells and macrophages across spatial scales is associated with survival.

The proposed methods are implemented in \texttt{R} and publicly available on \url{https://github.com/julia-wrobel/KAMP-paper}. Ripley's $K$ with a translation edge correction is implemented using the \texttt{spatstat} package \citep{baddeley2015spatial}. Our methods have also been incorporated as part of the \texttt{spatialTIME} and \texttt{mxfda} \texttt{R} packages \citep{creed2021spatialtime, wrobel2024mxfda}.


\subsection{Data cleaning}

The HGSOC dataset originates from a study of 128 subjects, collected at the University of Colorado and containing one sample per subject. Spatial proteomics images were acquired using the Vectra Automated Quantitative Pathology System (Akoya Biosciences). Image analysis, including tissue segmentation to define cells within tumor regions, cell segmentation, and cell phenotyping, was performed using inForm software. Detailed descriptions of the data and image pre-processing steps can be found in \citep{jordan2020capacity, steinhart2021spatial}. The dataset is publicly available in a tabular post-segmentation format through the Bioconductor package \texttt{VectraPolarisData}, and includes patient-level variables such as age, cancer stage at diagnosis, survival time, and survival status.

Our analysis focuses on a subset of patients with primary tumor samples, defined as those collected prior to clinical intervention. We focus exclusively on primary tumor samples for two key reasons. First, post-treatment data are unavailable, and including unmatched non-primary samples would obscure meaningful patterns due to substantial tumor heterogeneity \citep{lee2015tumor}. Second, treatment protocols varied across the 25 treated patients, introducing additional complexity. Limiting our analysis to primary samples enables a clearer interpretation of patient outcomes at initial presentation, aligning with standard practices in the field \citep{achimas2022evolutionary}.

Cancer stage was dichotomized into stages 1 and 2 versus stages 3 and 4, and the analysis was restricted to tumor tissue areas. The final analytic dataset for modeling \ clustering of all immune cells includes 103 subjects, with a median [IQR] of 10,373 [7,350, 13,156] total cells and 409 [159, 886] immune cells per sample, corresponding to a proportion of immune cells of 0.045 [0.016, 0.099].  For modeling spatial colocalization of B cells and macrophages, we restricted the analytic dataset to patients with at least two B cells and two macrophages each, resulting in a dataset of 71 subjects, with a median [IQR] of  9,383 [7,048, 11,452] total cells, 8 [4, 39] B cells and 262 [108, 612] macrophages.


\subsection{Within-sample analysis}

We estimate the degree of clustering $\tilde{K}(r)$ using \textit{K}, \textit{Kinhom}, \textit{KAMP}, \textit{KAMP lite}, and \textit{perm} with 10,000 permutations. We calculate $p$-values to evaluate significant clustering within each image across a range of radii, $r \in (0, 200)$, using \textit{KAMP}, \textit{KAMP lite}, and \textit{perm}. 

Computation times for our \textit{KAMP} and \textit{KAMP lite} approaches were substantially faster than performing 10,000 permutations for each image. Notably, for the entire dataset the \textit{perm} approach took 562 minutes whereas \textit{KAMP} and \textit{KAMP lite} took 3.6 and 0.6 minutes, respectively. On a per-image basis, median compute times were 213.1, 1.6, and 0.2 seconds for \textit{perm}, \textit{KAMP}, and \textit{KAMP lite}. 

\begin{figure}[h!]
	\centering
	\includegraphics[width=\columnwidth]{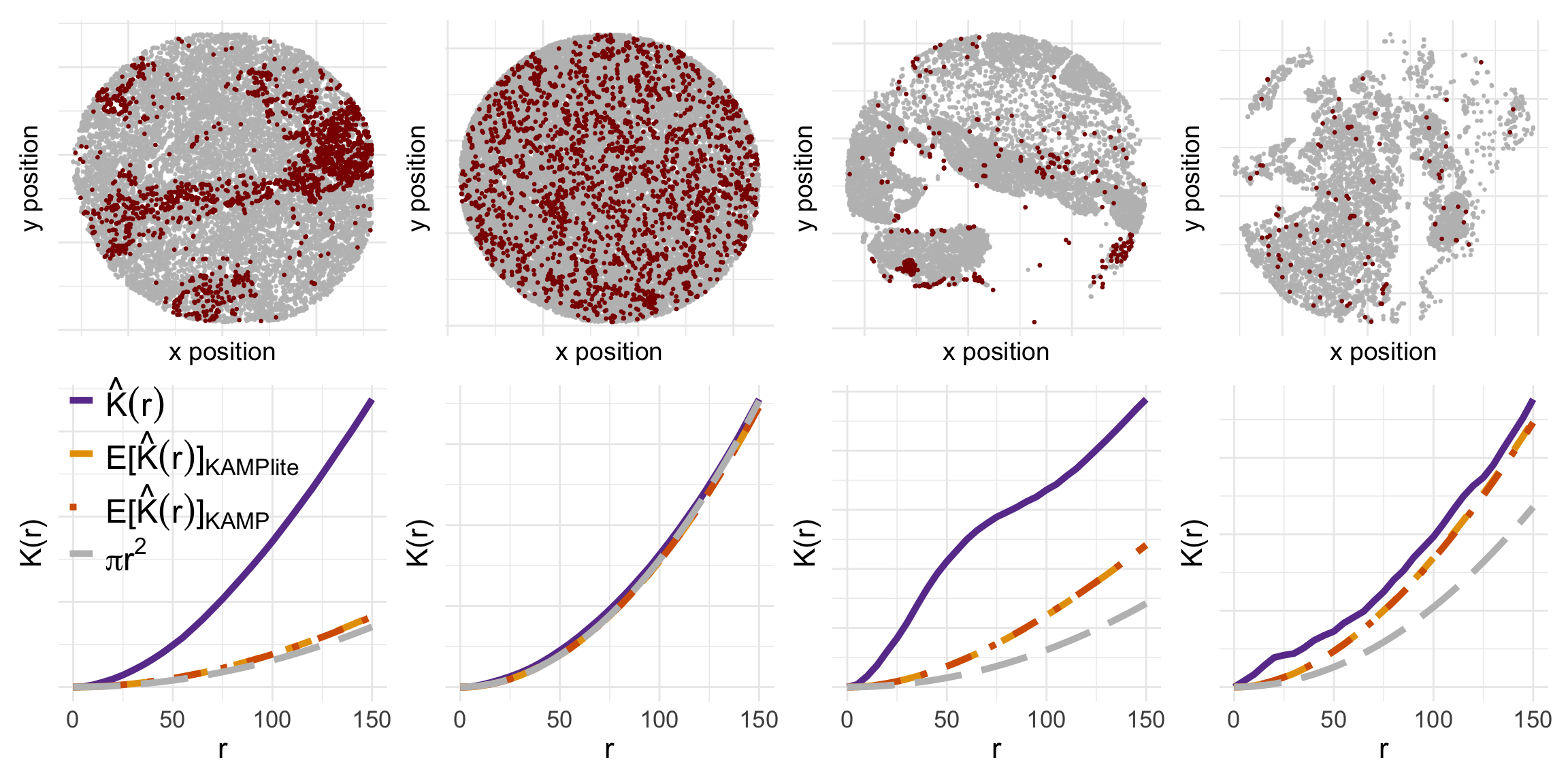}
	\caption{Observed point patterns and $K(r)$ estimates from the four HGSOC samples in Figure \ref{fig:patchy_images}. The bottom row shows $\hat{K}(r)$ (in purple), the theoretical null value under spatial homogeneity ($\pi r^2$, gray), and the empirical null estimated using the expectation of the permutation distribution under \textit{KAMP} (green) and \textit{KAMP lite} with 50\% thinning (yellow), denoted $E_{KAMP}[\hat{K}(r)]$ and $E_{KAMPlite}[\hat{K}(r)]$, respectively.}
	\label{fig:patchy_images_KAMP}
\end{figure}

After applying our method, images with technical inhomogeneity should have a \textit{KAMP} and \textit{KAMP lite}-estimated $K_0(r)$ value that is higher than the theoretical $K_0(r)$ value of $\pi r^2$. Figure \ref{fig:patchy_images_KAMP} illustrates this point by replotting subjects from Figure \ref{fig:patchy_images} with the inclusion of \textit{KAMP} and \textit{KAMP lite} $K_0(r)$ values. In particular the patient's sample in the third column of Figure \ref{fig:patchy_images_KAMP} displays clear spatial inhomogeneity due to tissue tearing, as well as apparent biological clustering of immune cells. 

The \textit{KAMP} $K_0(r)$ values for this sample (green dotted line) lie well above $\pi r^2$ (gray line) across $r$ but fall below the observed estimate $\hat{K}(r)$ (purple line). This suggests that our method effectively removes technical inhomogeneity while preserving signal from biological clustering. The patient's sample in the fourth column also shows inhomogeneity, but immune cells appear more dispersed.  Here, the \textit{KAMP} $K_0(r)$ values closely follow the observed estimate $\hat{K}(r)$, indicating that \textit{KAMP} corrects for inhomogeneity that \textit{K} mistakenly interprets as immune cell clustering. In contrast, for the patients in the left and middle columns, which do not appear to have substantial inhomogeneity, the \textit{KAMP} $K_0(r)$ values align closely with $\pi r^2$. Notably, $K_0(r)$ values for \textit{KAMP} and \textit{KAMP lite} are nearly identical for these four samples. 

\begin{figure}[h!]
	\centering
	\includegraphics[width=\columnwidth]{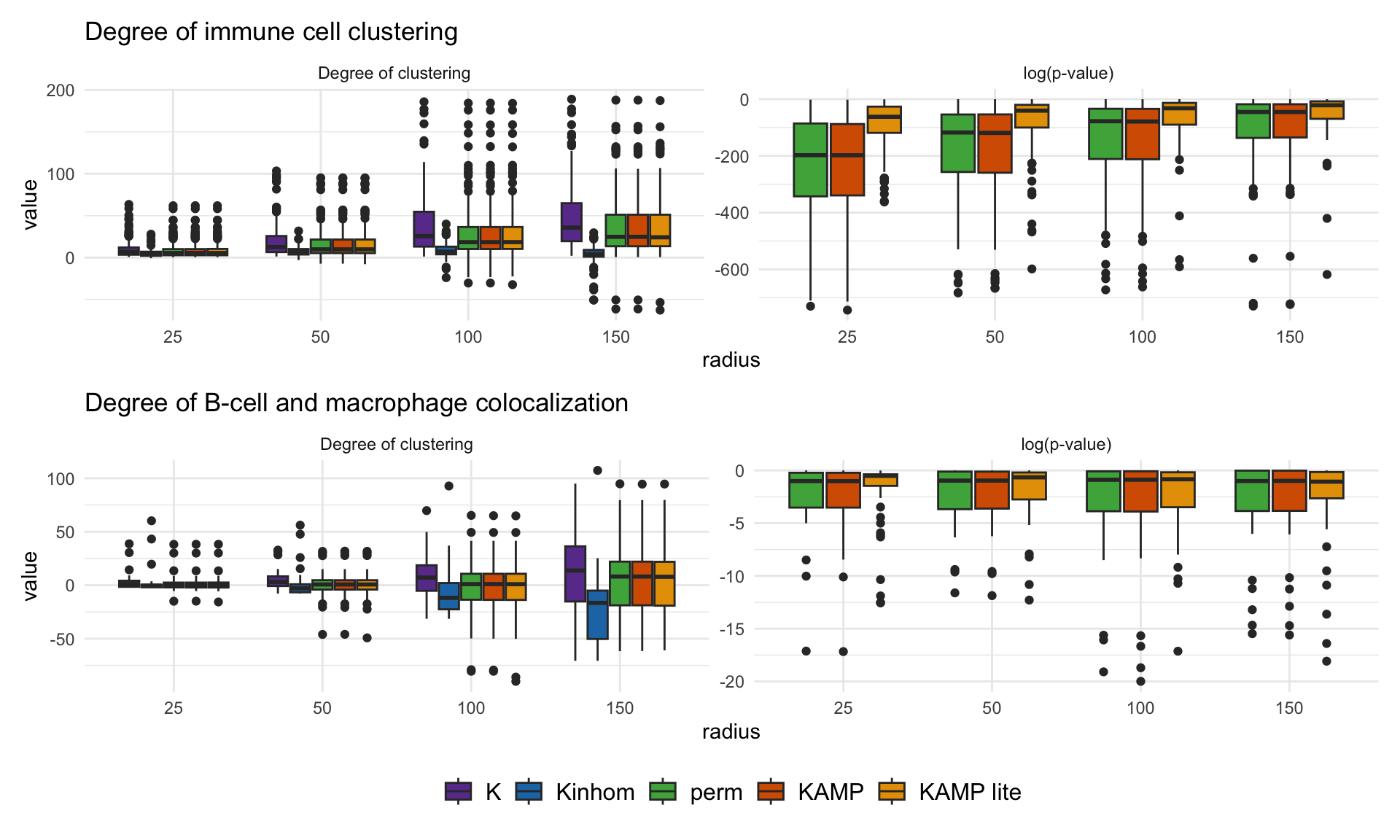}
	\caption{Boxplots comparing $\tilde{K}_s(r)$ (top left) and associated within-sample $p$-values testing positive excess clustering (top right), as well as the degree of B cell and macrophage colocalization $\tilde{K}_{sc}(r)$ (bottom left) and corresponding within-sample $p$-values (bottom right), across methods and radii $r \in \{25, 50, 100, 150\}$. \textit{KAMP lite} results use a thinning probability of 0.5.}
	\label{fig:ovarian_boxplots}
\end{figure}

Figure \ref{fig:ovarian_boxplots} presents boxplots comparing the degree of clustering (left panels) and the log within-sample $p$-values (right panels) across methods and radii $r \in \{25, 50, 100, 150\}$. The top row summarizes univariate clustering of all immune cells, while the bottom row shows bivariate colocalization of B cells and macrophages. In the top-left panel, the \textit{K} method consistently produces higher $\tilde{K}(r)$ estimates across radii, while the estimates from  \textit{KAMP}, \textit{KAMP lite}, and \textit{perm} are smaller and closely aligned. The \textit{Kinhom} method produces substantially lower estimates than all other methods. These findings are consistent with the simulation results in Section \ref{sec:simulations}, as several samples in the dataset exhibit inhomogeneity. Because \textit{K} does not account for inhomogeneity, it tends to inflate $\tilde{K}(r)$, whereas the \textit{KAMP}, \textit{KAMP lite}, and \textit{perm} methods appropriately adjust for it.

The top-right panel displays log $p$-values for the one-sided within-sample test of positive excess clustering relative to the permutation-null benchmark. As \textit{K} and \textit{Kinhom} do not support within-sample inference, no $p$-values are shown for these methods. \textit{KAMP} and \textit{perm} produce similar $p$-value distributions, as expected since \textit{perm}  approximates the full permutation distribution analytically computed by \textit{KAMP}. All methods show smaller $p$-values at smaller radii indicating more significant spatial clustering at smaller distances, but \textit{KAMP lite} tends to produce more conservative (i.e., larger) $p$-values due to increased variance introduced by thinning.

The bottom row mirrors these comparisons for B cell and macrophage colocalization. Again, the \textit{K} method yields inflated $\tilde{K}_{sc}(r)$ values, reflecting its sensitivity to spatial inhomogeneity. As with the univariate clustering analysis, \textit{KAMP} and \textit{KAMP lite} produce similar colocalization estimates, while \textit{KAMP lite} yields more conservative $p$-values. These results reinforce that while \textit{KAMP lite} may be useful for estimating degree-of-clustering metrics, \textit{KAMP} is better suited for reliable within-sample inference. 


\subsection{Across-sample analysis} \label{subsec:across}

In the across‑sample analysis, we examine spatial colocalization of B cells and macrophages within the tumor compartment and its association with patient survival using Cox proportional hazards models. Each model includes $\tilde{K}_{sc}(r)$ at radius $r \in \{5, 10, \ldots, 200\}$ as a covariate, along with age, cancer stage, and immune-cell proportion. Age, cancer stage, and immune-cell proportion are included as adjustment covariates rather than as primary targets of inference. We fit separate models using $\tilde{K}_{sc}(r)$ estimated by the \textit{K}, \textit{Kinhom}, \textit{perm}, \textit{KAMP}, and \textit{KAMP lite} methods, and compare results based on standardized effect sizes, Cox-model $p$-values for the spatial colocalization coefficient, and concordance indices \citep{vandekar2020robust, jones2023resi}. The radii are analyzed as a fixed exploratory grid. The $p$-values and confidence intervals reported below are pointwise and are not adjusted for multiple testing across radii.

\begin{figure}[h!]
	\centering
	\includegraphics[width=\columnwidth]{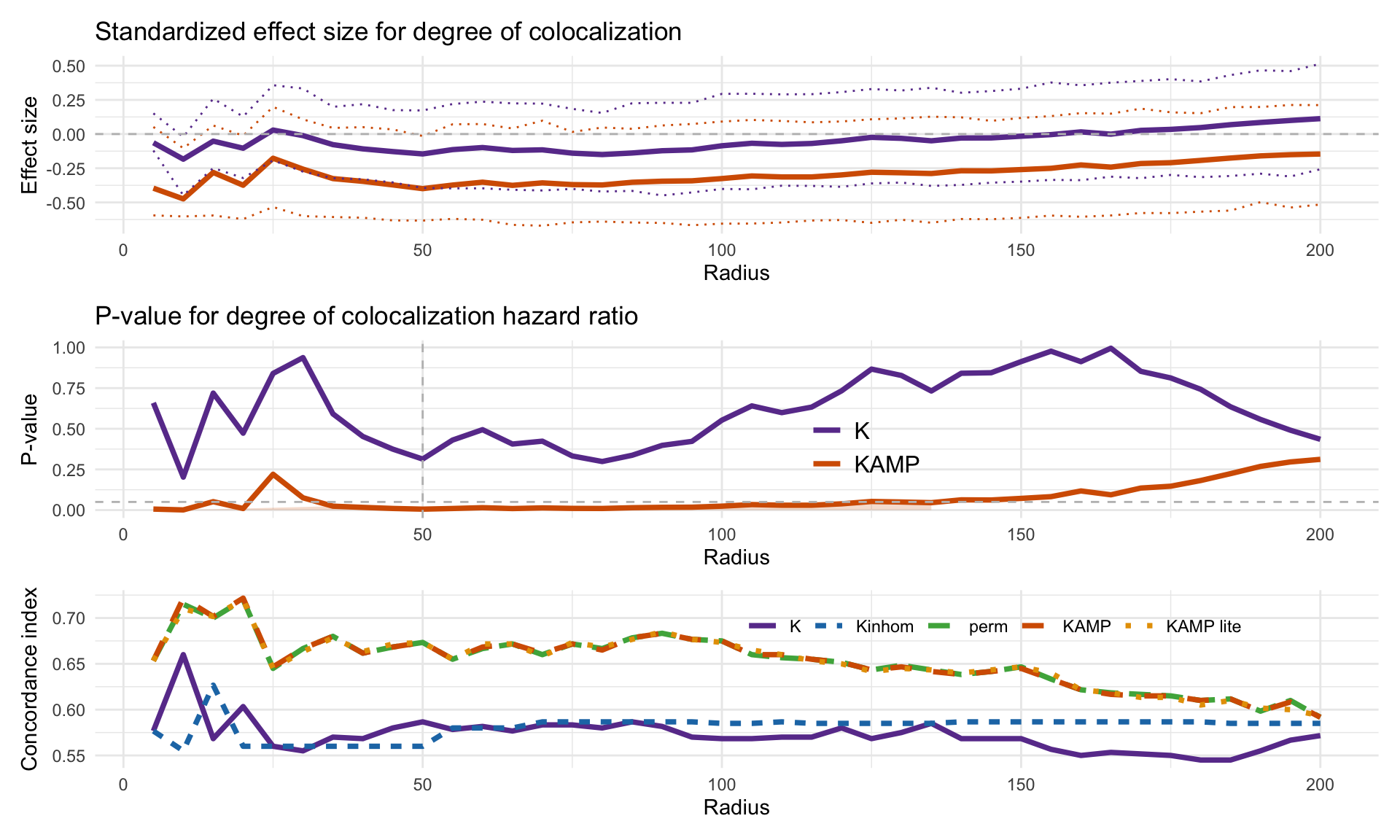}
	\caption{$\tilde{K}_{sc}(r)$ modeling results for B cells and macrophages in relation to overall survival. The top panel shows signed RESI estimates with pointwise 95\% nonparametric patient-bootstrap intervals for \textit{K} and \textit{KAMP}. The second panel displays Cox-model $p$-values for testing the spatial colocalization coefficient. The bottom panel shows the c-index across radii for each method, where higher values indicate better in-sample discrimination. \textit{KAMP lite} results use a thinning probability of 0.5.}
	\label{fig:ovarian_survival}   
\end{figure}

Figure \ref{fig:ovarian_survival} presents these comparisons across radii. The top panel shows standardized effect sizes (solid lines) and bootstrapped 95\% confidence intervals (dotted lines) for the \textit{K} and \textit{KAMP} methods. The confidence intervals are obtained using a nonparametric bootstrap with 1,000 bootstrap samples where patients are resampled with replacement \citep{vandekar2020robust}. The second panel displays nominal pointwise Cox $p$-values for the spatial colocalization coefficient. The bottom panel shows concordance index (c-index) values across radii and methods, where higher values reflect better predictive performance. 

Models using \textit{K}, which does not adjust for spatial inhomogeneity, show smaller standardized effect sizes, higher $p$-values, and lower c-index values than those using \textit{KAMP}. The top panel of Figure \ref{fig:ovarian_survival} shows that \textit{KAMP} consistently estimates a stronger protective association between B cell/macrophage colocalization and survival across nearly the entire range of radii considered. The second panel shows that, at the nominal pointwise level $\alpha=0.05$, the \textit{KAMP}-derived degree of colocalization has $p$-values below 0.05 at $r=5, 10, 20$, at every 5-unit radius from $r=35$ through $r=120$, and at $r=130$ and $r=135$, whereas $K$ shows no statistically significant associations at any radius. Finally, the bottom panel shows that models using \textit{KAMP} consistently achieve higher c-index values than those using \textit{K}, indicating improved in-sample discrimination. Together, these results suggest that inhomogeneity bias in $\tilde{K}_{sc}(r)$ compromises its reliability for survival modeling, reinforcing the advantages of \textit{KAMP}. We note that age, cancer stage, and immune-cell proportion are included as adjustment covariates rather than as primary targets of inference. Given the modest analytic sample size ($n = 71$) and the limited variability of the dichotomized stage variable, we do not draw substantive conclusions from their individual coefficients further.

As an exploratory analysis, we also evaluate the association between overall immune-cell clustering and overall survival using the univariate \textit{KAMP} statistic; results are presented in Supplement F.


\section{Discussion} \label{sec:discussion}

In this work, we introduce \textit{KAMP}, a novel method for quantifying immune cell clustering and colocalization in spatial proteomics data that accounts for spatial inhomogeneity. \textit{KAMP} provides both spatial summary metrics for downstream modeling and sample-level hypothesis testing. By deriving analytical moments of the permutation distribution of Ripley's $K$, our approach corrects for bias introduced by tissue degradation and other artifacts, a common challenge in spatial proteomics. Crucially, \textit{KAMP} achieves substantial computational efficiency over permutation-based methods without sacrificing accuracy, power, or Type I error, making it well-suited for the analysis of increasingly large spatial datasets.

Applied to HGSOC, \textit{KAMP} and conventional \textit{K} yield materially different radius-specific estimated associations between B cell–macrophage colocalization and overall survival. The \textit{KAMP} estimates are generally more negative and the \textit{KAMP}-based models has higher apparent in-sample concordance, although the pointwise uncertainty in the effect-size estimates
is substantial. Notably, \textit{KAMP} is over 100 times faster than explicit permutation in this dataset. Although our motivating application focuses on B cell–macrophage colocalization, the proposed bivariate \textit{KAMP} framework is applicable to arbitrary pairs of cell types. We anticipate that the method will be useful for investigating a broad range of biologically motivated cell-cell interactions in future spatial proteomics studies. 

As an exploratory supplementary analysis, we also apply the univariate \textit{KAMP} statistic to overall immune-cell clustering in the primary tumor samples used for the univariate analysis $(n = 103)$. This analysis shows positive Cox coefficients across the radius range and nominal pointwise associations with overall survival at intermediate radii (Supplement F). However, these associations are weaker and less consistent than the prespecified B cell--macrophage colocalization result emphasized in the main application. The univariate clustering statistic also has a different biological interpretation: it summarizes the overall spatial aggregation of immune cells and may reflect broad tumor-immune architecture, whereas the B cell--macrophage analysis targets a specific cell-cell colocalization hypothesis motivated by ovarian cancer biology.

One important implication of the \textit{KAMP} estimand is that it measures excess cell-type-specific clustering relative to the pooled tissue structure, rather than absolute clustering relative to complete spatial randomness. Therefore, when the biological signal of interest is shared by immune and background cells or is strongly aligned with the underlying tissue architecture, \textit{KAMP} and other permutation-null approaches may remove much of this shared spatial structure. In such settings, \textit{KAMP} may have reduced power for detecting residual immune-specific clustering, and ordinary Ripley’s $K$ or other summaries of absolute spatial aggregation may be more appropriate if the scientific target is total clustering relative to complete spatial randomness.

It is also important to note that the choice of radius $r$ determines the spatial scale over which clustering or colocalization is quantified. When possible, we recommend pre-specifying a small number of radii based on prior biological knowledge of the expected interaction distance between the cell types of interest. In practice, however, such interaction distances are often unknown for emerging spatial proteomics technologies. In these settings, one could instead consider penalized regression approaches that select among a larger set of candidate radii. Another approach is to model the entire $\tilde{K}_{sc}(r)$ curve rather than fitting separate models at each radius. In this work, we considered a sequence of fixed radii to facilitate interpretation and compatibility with standard survival modeling approaches. Future work could instead treat $\tilde{K}_{sc}(r)$ as a functional covariate in a functional Cox model, enabling inference on the complete spatial clustering trajectory while accounting for the correlation across radii \citep{wrobel2024mxfda}. 

Our method opens important directions for future research. First, although \textit{KAMP} is efficient for datasets with roughly 10,000 cells per sample, newer technologies now produce images with over a million cells, necessitating further computational improvements. Our lightweight variant, \textit{KAMP lite}, offers substantial speedups and performs well for estimating expected clustering for use in cross-sample modeling. However, for within-sample hypothesis testing, we recommend the full \textit{KAMP} procedure due to its more accurate variance estimates and improved power.

Another important direction is to propagate uncertainty in \textit{KAMP}-derived summaries into downstream patient-level models. In this work, the estimated degree of clustering or colocalization, $\tilde{K}_{s}(r)$ and $\tilde{K}_{sc}(r)$, is treated as a fixed covariate in the Cox proportional hazards models after it is computed from the observed cell locations and labels. Although the analytic permutation moments eliminate Monte Carlo uncertainty from estimating the permutation-null benchmark, the resulting summaries may still reflect uncertainty due to finite cell counts, segmentation and phenotyping errors, and the choice of radius. Future work could incorporate this uncertainty using measurement-error models, bootstrap or resampling procedures over cells or tissue regions, or hierarchical models that jointly estimate spatial summaries and their association with clinical outcomes.

Second, as with any method relying on cell-type annotation, \textit{KAMP}'s validity depends on accurate labeling. Errors from segmentation or ambiguous marker expression can distort spatial structure and bias inference. Since \textit{KAMP} does not explicitly correct for label noise, its performance hinges on high-quality phenotyping. Future work could incorporate probabilistic or uncertainty-aware labeling to improve robustness.

Third, while the current \textit{KAMP} framework focuses on univariate and pairwise interactions, it complements emerging approaches that analyze multitype spatial structure using entropy or topological data analysis \citep{vu2023funspace, samorodnitsky2024detecting}. These methods summarize global spatial organization but do not isolate specific cell-type interactions. Extending \textit{KAMP} to detect higher-order colocalization patterns is a promising avenue.

Another useful direction is to extend the analytic permutation-moment framework beyond Ripley's $K$. For example, Moran's $I$ \citep{moran1950} and other spatial association statistics can also be formulated using cell labels and spatial weights, suggesting that analogous empirical null approximations may be possible. Such extensions would require statistic-specific choices, such as the spatial weight matrix for Moran's $I$, and would address related but distinct scientific questions from the radius-dependent clustering and colocalization analyses considered here.

Finally, although our application centers on spatial proteomics, the \textit{KAMP} framework is broadly applicable to other domains with repeated multitype point patterns, including satellite imaging, ecology, and spatial transcriptomics.  Along these lines, other recent work has proposed similar analytical approaches for constructing empirical null distributions for spatial statistics used in spatial omics research \citep{hawinkel2025unified}.


\section{Significance statement}

Spatial proteomics reveals where different cell types are located within tissue, allowing researchers to study how immune-cell organization relates to patient outcomes. However, uneven cell density caused by tissue architecture or sample damage can be mistaken for clustering of particular cell types. Our new approach \textit{KAMP} addresses this problem by using the spatial arrangement of all observed cells as a reference to quantify clustering and colocalization beyond the background tissue structure. Its key innovation is to replace repeated random reshuffling of cell labels with direct formulas for the expected level of clustering and its variability, substantially reducing computational cost. In an exploratory analysis of ovarian cancer samples, greater colocalization of B cells and macrophages, two types of immune cells, was associated with better overall survival at several spatial scales. These associations were not detected using the unadjusted method. Our findings highlight the importance of accounting for background tissue structure when relating spatial cell patterns to clinical outcomes and provide a practical tool for investigating immune-cell organization in large spatial proteomics datasets.


\begin{acks}[Acknowledgments]
	Julia Wrobel and Hoseung Song contributed equally to this work. Hoseung Song is supported, in part, by the National Research Foundation of Korea(NRF) grant funded by the Korea government(MSIT) (RS-2022-NR068758,RS-2026-25471551). We thank the Editor, the Associate Editor, and the two anonymous reviewers for their thoughtful and constructive comments, which substantially improved the manuscript. We particularly thank one of the reviewers for suggesting the sparse, KD-tree-based computational improvements described in Section 2.3.
\end{acks}


\begin{supplement}
Supplementary material contains proofs for Theorem 2.1 and 2.4, simulation results for bivariate \textit{KAMP} and \textit{KAMP lite}, additional simulations in which both background and immune cells exhibit spatial clustering, and exploratory survival analysis of overall immune-cell clustering.
\end{supplement}


\bibliographystyle{imsart-nameyear} 
\bibliography{biblio}       

@article{moran1950,
  author = {Moran, P. A. P.},
  title = {Notes on Continuous Stochastic Phenomena},
  journal = {Biometrika},
  volume = {37},
  number = {1--2},
  pages = {17--23},
  year = {1950},
  doi = {10.1093/biomet/37.1-2.17}
}

@book{ripley1988statistical,
	title={Statistical inference for spatial processes},
	author={Ripley, Brian D},
	year={1988},
	publisher={Cambridge university press}
}

@book{diggle2013statistical,
	title={Statistical analysis of spatial and spatio-temporal point patterns},
	author={Diggle, Peter J},
	year={2013},
	publisher={CRC press}
}

@article{palla2022squidpy,
	title={Squidpy: a scalable framework for spatial omics analysis},
	author={Palla, Giovanni and Spitzer, Hannah and Klein, Michal and Fischer, David and Schaar, Anna Christina and Kuemmerle, Louis Benedikt and Rybakov, Sergei and Ibarra, Ignacio L and Holmberg, Olle and Virshup, Isaac and others},
	journal={Nature methods},
	volume={19},
	number={2},
	pages={171--178},
	year={2022},
	publisher={Nature Publishing Group US New York}
}

@inproceedings{lagache2013statistical,
  title={A statistical analysis of spatial colocalization using Ripley's K function},
  author={Lagache, Thibault and Meas-Yedid, Vannary and Olivo-Marin, Jean-Christophe},
  booktitle={2013 IEEE 10th International Symposium on Biomedical Imaging},
  pages={896--901},
  year={2013},
  organization={IEEE}
}

@article{gupta2019b,
  title={B cells as an immune-regulatory signature in ovarian cancer},
  author={Gupta, Prachi and Chen, Changliang and Chaluvally-Raghavan, Pradeep and Pradeep, Sunila},
  journal={Cancers},
  volume={11},
  number={7},
  pages={894},
  year={2019},
  publisher={MDPI}
}

@article{lan2013expression,
  title={Expression of M2-polarized macrophages is associated with poor prognosis for advanced epithelial ovarian cancer},
  author={Lan, Chunyan and Huang, Xin and Lin, Suxia and Huang, Huiqiang and Cai, Qichun and Wan, Ting and Lu, Jiabin and Liu, Jihong},
  journal={Technology in cancer research \& treatment},
  volume={12},
  number={3},
  pages={259--267},
  year={2013},
  publisher={SAGE Publications Sage CA: Los Angeles, CA}
}

@article{nowak2020role,
  title={The role of tumor-associated macrophages in the progression and chemoresistance of ovarian cancer},
  author={Nowak, Marek and Klink, Magdalena},
  journal={Cells},
  volume={9},
  number={5},
  pages={1299},
  year={2020},
  publisher={MDPI}
}

@article{song2022fast,
  title={A fast kernel independence test for cluster-correlated data},
  author={Song, Hoseung and Liu, Hongjiao and Wu, Michael C},
  journal={Scientific Reports},
  volume={12},
  number={1},
  pages={21659},
  year={2022},
  publisher={Nature Publishing Group UK London}
}

@article{lee2015tumor,
  title={Tumor evolution and intratumor heterogeneity of an epithelial ovarian cancer investigated using next-generation sequencing},
  author={Lee, Jung-Yun and Yoon, Jung-Ki and Kim, Boyun and Kim, Soochi and Kim, Min A and Lim, Hyeonseob and Bang, Duhee and Song, Yong-Sang},
  journal={BMC cancer},
  volume={15},
  pages={1--9},
  year={2015},
  publisher={Springer}
}

@article{achimas2022evolutionary,
  title={Evolutionary perspectives, heterogeneity and ovarian cancer: a complicated tale from past to present},
  author={Achimas-Cadariu, Patriciu and Kubelac, Paul and Irimie, Alexandru and Berindan-Neagoe, Ioana and R{\"u}hli, Frank},
  journal={Journal of ovarian research},
  volume={15},
  number={1},
  pages={67},
  year={2022},
  publisher={Springer}
}

@article{liu2021kernel,
  title={A kernel-based test of independence for cluster-correlated data},
  author={Liu, Hongjiao and Plantinga, Anna and Xiang, Yunhua and Wu, Michael},
  journal={Advances in neural information processing systems},
  volume={34},
  pages={9869--9881},
  year={2021}
}

@article{davies1980distribution,
  title={The distribution of a linear combination of $\chi$2 random variables},
  author={Davies, Robert B},
  journal={Journal of the Royal Statistical Society Series C: Applied Statistics},
  volume={29},
  number={3},
  pages={323--333},
  year={1980},
  publisher={Oxford University Press}
}

@article{zhan2017fast,
  title={A fast small-sample kernel independence test for microbiome community-level association analysis},
  author={Zhan, Xiang and Plantinga, Anna and Zhao, Ni and Wu, Michael C},
  journal={Biometrics},
  volume={73},
  number={4},
  pages={1453--1463},
  year={2017},
  publisher={Oxford University Press}
}

@article{mielke1984,
  title={Meteorological applications of permutation techniques based on distance functions},
  author={Mielke Jr, Paul W},
  journal={Handbook of statistics},
  volume={4},
  pages={813--830},
  year={1984},
  publisher={Elsevier}
}

@article{heo1998permutation,
  title={A permutation test of association between configurations by means of the RV coefficient},
  author={Heo, Moonseong and Ruben Gabriel, K},
  journal={Communications in Statistics-Simulation and Computation},
  volume={27},
  number={3},
  pages={843--856},
  year={1998},
  publisher={Taylor \& Francis}
}

@article{abdi2007rv,
  title={RV coefficient and congruence coefficient},
  author={Abdi, Herv{\'e}},
  journal={Encyclopedia of measurement and statistics},
  volume={849},
  number={853},
  pages={92},
  year={2007}
}

@article{bressan2023dawn,
  title={The dawn of spatial omics},
  author={Bressan, Dario and Battistoni, Giorgia and Hannon, Gregory J},
  journal={Science},
  volume={381},
  number={6657},
  pages={eabq4964},
  year={2023},
  publisher={American Association for the Advancement of Science}
}

@article{baddeley2000non,
  title={Non-and semi-parametric estimation of interaction in inhomogeneous point patterns},
  author={Baddeley, Adrian J and M{\o}ller, Jesper and Waagepetersen, Rasmus},
  journal={Statistica Neerlandica},
  volume={54},
  number={3},
  pages={329--350},
  year={2000},
  publisher={Wiley Online Library}
}

@book{baddeley2015spatial,
  title={Spatial point patterns: methodology and applications with R},
  author={Baddeley, Adrian and Rubak, Ege and Turner, Rolf},
  year={2015},
  publisher={CRC press}
}

@article{baddeley2005spatstat,
  title={Spatstat: an R package for analyzing spatial point patterns},
  author={Baddeley, Adrian and Turner, Rolf},
  journal={Journal of statistical software},
  volume={12},
  pages={1--42},
  year={2005}
}

@article{creed2021spatialtime,
  title={spatialTIME and iTIME: R package and Shiny application for visualization and analysis of immunofluorescence data},
  author={Creed, Jordan H and Wilson, Christopher M and Soupir, Alex C and Colin-Leitzinger, Christelle M and Kimmel, Gregory J and Ospina, Oscar E and Chakiryan, Nicholas H and Markowitz, Joseph and Peres, Lauren C and Coghill, Anna and others},
  journal={Bioinformatics},
  volume={37},
  number={23},
  pages={4584--4586},
  year={2021},
  publisher={Oxford University Press}
}

@article{jones2023resi,
  title={RESI: An R Package for Robust Effect Sizes},
  author={Jones, Megan and Kang, Kaidi and Vandekar, Simon},
  journal={arXiv preprint arXiv:2302.12345},
  year={2023}
}

@article{jordan2020capacity,
  title={The capacity of the ovarian cancer tumor microenvironment to integrate inflammation signaling conveys a shorter disease-free interval},
  author={Jordan, Kimberly R and Sikora, Matthew J and Slansky, Jill E and Minic, Angela and Richer, Jennifer K and Moroney, Marisa R and Hu, Junxiao and Wolsky, Rebecca J and Watson, Zachary L and Yamamoto, Tomomi M and Costello, James C and Clauset, Aaron and Behbakht, Kian and Kumar, T Rajendra and Bitler, Ben G},
  journal={Clinical Cancer Research},
  volume={26},
  number={23},
  pages={6362--6373},
  year={2020},
  publisher={AACR}
}

@article{shaw2021globally,
  title={Globally intensity-reweighted estimators for K-and pair correlation functions},
  author={Shaw, Thomas and Moller, Jesper and Waagepetersen, Rasmus Plenge},
  journal={Australian \& New Zealand Journal of Statistics},
  volume={63},
  number={1},
  pages={93--118},
  year={2021},
  publisher={Wiley Online Library}
}

@article{hawinkel2025unified,
  title={Unified nonparametric analysis of single-molecule spatial omics data using probabilistic indices},
  author={Hawinkel, Stijn and Yang, Xilan and Poelmans, Ward and Motte, Hans and Beeckman, Tom and Maere, Steven},
  journal={bioRxiv},
  pages={2025--05},
  year={2025},
  publisher={Cold Spring Harbor Laboratory}
}

@article{siegel2024cancer,
  title={Cancer statistics, 2024.},
  author={Siegel, Rebecca L and Giaquinto, Angela N and Jemal, Ahmedin},
  journal={CA: a cancer journal for clinicians},
  volume={74},
  number={1},
  year={2024}
}

@article{soupir2025scspatialsim,
  title={scSpatialSIM: a simulator of spatial single-cell molecular data},
  author={Soupir, Alex et al.},
  journal={SoftwareX},
  volume={31},
  pages={102223},
  year={2025},
  publisher={Elsevier}
}

@article{samorodnitsky2024detecting,
  title={Detecting Clinically Relevant Topological Structures in Multiplexed Spatial Proteomics Imaging Using TopKAT},
  author={Samorodnitsky, Sarah and Campbell, Katie and Little, Amarise and Ling, Wodan and Zhao, Ni and Chen, Yen-Chi and Wu, Michael C},
  journal={bioRxiv},
  year={2024}
}

@article{vu2023funspace,
  title={FunSpace: a functional and spatial analytic approach to cell imaging data using entropy measures},
  author={Vu, Thao and Seal, Souvik and Ghosh, Tusharkanti and Ahmadian, Mansooreh and Wrobel, Julia and Ghosh, Debashis},
  journal={PLOS Computational Biology},
  volume={19},
  number={9},
  pages={e1011490},
  year={2023},
  publisher={Public Library of Science San Francisco, CA USA}
}

@article{steinhart2021spatial,
  title={The spatial context of tumor-infiltrating immune cells associates with improved ovarian cancer survival},
  author={Steinhart, Benjamin and Jordan, Kimberly R and Bapat, Jaidev and Post, Miriam D and Brubaker, Lindsay W and Bitler, Benjamin G and Wrobel, Julia},
  journal={Molecular Cancer Research},
  volume={19},
  number={12},
  pages={1973--1979},
  year={2021},
  publisher={AACR}
}

@article{vandekar2020robust,
  title={A robust effect size index},
  author={Vandekar, Simon and Tao, Ran and Blume, Jeffrey},
  journal={Psychometrika},
  volume={85},
  number={1},
  pages={232--246},
  year={2020},
  publisher={Springer}
}

@article{wilson2021challenges,
  title={Challenges and opportunities in the statistical analysis of multiplex immunofluorescence data},
  author={Wilson, Christopher M and Ospina, Oscar E and Townsend, Mary K and Nguyen, Jonathan and Moran Segura, Carlos and Schildkraut, Joellen M and Tworoger, Shelley S and Peres, Lauren C and Fridley, Brooke L},
  journal={Cancers},
  volume={13},
  number={12},
  pages={3031},
  year={2021},
  publisher={MDPI}
}

@article{wilson2022tumor,
  title={Tumor immune cell clustering and its association with survival in African American women with ovarian cancer},
  author={Wilson, Christopher and Soupir, Alex C and Thapa, Ram and Creed, Jordan and Nguyen, Jonathan and Segura, Carlos Moran and Gerke, Travis and Schildkraut, Joellen M and Peres, Lauren C and Fridley, Brooke L},
  journal={PLoS computational biology},
  volume={18},
  number={3},
  pages={e1009900},
  year={2022},
  publisher={Public Library of Science San Francisco, CA USA}
}

@article{wrobel2023statistical,
  title={Statistical analysis of multiplex immunofluorescence and immunohistochemistry imaging data},
  author={Wrobel, Julia and Harris, Coleman and Vandekar, Simon},
  journal={Statistical Genomics},
  pages={141--168},
  year={2023},
  publisher={Springer}
}

@article{winkler2014permutation,
  title={Permutation inference for the general linear model},
  author={Winkler, Anderson M and Ridgway, Gerard R and Webster, Matthew A and Smith, Stephen M and Nichols, Thomas E},
  journal={Neuroimage},
  volume={92},
  pages={381--397},
  year={2014},
  publisher={Elsevier}
}

@article{keren2018structured,
  title={A structured tumor-immune microenvironment in triple negative breast cancer revealed by multiplexed ion beam imaging},
  author={Keren, Leeat and Bosse, Marc and Marquez, Diana and Angoshtari, Roshan and Jain, Samir and Varma, Sushama and Yang, Soo-Ryum and Kurian, Allison and Van Valen, David and West, Robert and others},
  journal={Cell},
  volume={174},
  number={6},
  pages={1373--1387},
  year={2018},
  publisher={Elsevier}
}

@article{hahn2012studentized,
  title={A studentized permutation test for the comparison of spatial point patterns},
  author={Hahn, Ute},
  journal={Journal of the American Statistical Association},
  volume={107},
  number={498},
  pages={754--764},
  year={2012},
  publisher={Taylor \& Francis}
}

@article{shinohara2020distance,
  title={Distance-based analysis of variance for brain connectivity},
  author={Shinohara, Russell T and Shou, Haochang and Carone, Marco and Schultz, Robert and Tunc, Birkan and Parker, Drew and Martin, Melissa Lynne and Verma, Ragini},
  journal={Biometrics},
  volume={76},
  number={1},
  pages={257--269},
  year={2020},
  publisher={Oxford University Press}
}

@article{minas2014distance,
  title={Distance-based analysis of variance: approximate inference},
  author={Minas, Christopher and Montana, Giovanni},
  journal={Statistical Analysis and Data Mining: The ASA Data Science Journal},
  volume={7},
  number={6},
  pages={450--470},
  year={2014},
  publisher={Wiley Online Library}
}

@article{vandereyken2023methods,
  title={Methods and applications for single-cell and spatial multi-omics},
  author={Vandereyken, Katy and Sifrim, Alejandro and Thienpont, Bernard and Voet, Thierry},
  journal={Nature Reviews Genetics},
  volume={24},
  number={8},
  pages={494--515},
  year={2023},
  publisher={Nature Publishing Group UK London}
}

@article{vu2022spf,
  title={SPF: a spatial and functional data analytic approach to cell imaging data},
  author={Vu, Thao and Wrobel, Julia and Bitler, Benjamin G and Schenk, Erin L and Jordan, Kimberly R and Ghosh, Debashis},
  journal={PLoS computational biology},
  volume={18},
  number={6},
  pages={e1009486},
  year={2022},
  publisher={Public Library of Science San Francisco, CA USA}
}

@article{wrobel2024mxfda,
  title={mxfda: a comprehensive toolkit for functional data analysis of single-cell spatial data},
  author={Wrobel, Julia and Soupir, Alex C and Hayes, Mitchell T and Peres, Lauren C and Vu, Thao and Leroux, Andrew and Fridley, Brooke L},
  journal={Bioinformatics Advances},
  volume={4},
  number={1},
  pages={vbae155},
  year={2024},
  publisher={Oxford University Press}
}

\end{document}